\documentclass[11pt,a4paper]{article}
\usepackage{jheppub}
\usepackage[utf8]{inputenc}
\usepackage[english]{babel}
\usepackage{color}
\usepackage{amsmath}
\usepackage{amsfonts}
\usepackage{amssymb}
\usepackage{graphicx}
\usepackage{orcidlink}
\usepackage{sidecap}
\usepackage[pdf]{pstricks} 
\usepackage{hyperref}
\usepackage{breakurl}
\hypersetup{breaklinks=true}
\newcommand{\change}[1]{\textcolor{black}{#1}}
\newcommand{\dx}{\mathrm{d}}
\newcommand{\eps}{\varepsilon}
\newcommand{\e}{\mathrm{e}}
\newcommand{\iu}{\mathrm{i}}
\newcommand{\dilog}{\mathrm{Li}_2}

\newcommand{\clausen}{\mathrm{Cl}_2}
\newcommand{\catalan}{\mathrm{C}}
\newcommand{\gieseking}{\mathrm{G}}
\newcommand{\vol}{\mathbb{V}}
\newcommand{\ghy}{{}_2\mathrm{F}_1}
\newcommand{\digam}{\psi^{(0)}}
\newcommand{\gammaeuler}{\gamma_\text{E}}
\sidecaptionvpos{figure}{c}
\begin{document}

\preprint{arXiv:2410.18177}

\title{Angular integrals with three denominators via IBP, mass reduction, dimensional shift, and differential equations}

\author{Juliane Haug and}
\emailAdd{juliane-clara-celine.haug@uni-tuebingen.de}%\,\orcidlink{0009-0008-3076-1122}
\author{Fabian Wunder}%\,\orcidlink{0009-0007-4136-7844}
\emailAdd{fabian.wunder@uni-tuebingen.de}
%\thanks{}
\affiliation{Institute for Theoretical Physics, University of T\"ubingen\\
Auf der Morgenstelle 14, 72076 T\"ubingen, Germany}

\date{October 25, 2024}

\abstract{
Angular integrals arise in a wide range of perturbative quantum field theory calculations.
In this work we investigate angular integrals with three denominators in $d=4-2\varepsilon$ dimensions.
We derive integration-by-parts relations for this class of integrals, leading to explicit recursion relations and a reduction to a small set of master integrals.
Using a differential equation approach we establish results up to order $\varepsilon$ for general integer exponents and masses.
Here, reduction identities for the number of masses, known results for two-denominator integrals, and a general dimensional-shift identity for angular integrals considerably reduce the required amount of work.
For the first time we find for angular integrals a term contributing proportional to a Euclidean Gram determinant in the $\varepsilon$-expansion.
This coefficient is expressed as a sum of Clausen functions with intriguing connections to Euclidean, spherical, and hyperbolic geometry.
The results of this manuscript are applicable to phase-space calculations with multiple observed final-state particles.}


\maketitle
%\tableofcontents
\flushbottom
 
\section{Introduction}
\label{sec: Introduction}
Angular integrals \cite{Schellekens:1981,vanNeerven:1985,Beenakker:1988,Somogyi:2011,Lyubovitskij:2021,Wunder:2024,Smirnov:2024} are an elementary building block of phase-space integrals arising in perturbative quantum field theory \cite{Bolzoni:2010,Anastasiou:2013, Lillard:2016,Kotlarski:2016,Lionetti:2018, Specchia:2018,Bahjat-Abbas:2018, Baranowski:2020,Blumlein:2020,Isidori:2020,Alioli:2022,Assi:2023,Catani:2023,Pal:2023,Devoto:2024,Rowe:2024}.
In the Quantum Chromodynamics (QCD) literature they have been used for example in theoretical predictions for the Drell-Yan process (DY) \cite{Matsuura:1989,Matsuura:1990,Hamberg:1991,Mirkes:1992,Bahjat-Abbas:2018}, deep-inelastic scattering (DIS) \cite{Duke:1982,Hekhorn:2019}, semi-inclusive deep-inelastic scattering (SIDIS) \cite{Anderle:2016,Wang:2019}, prompt-photon production \cite{Gordon:1993, Coriano:1996, Rein:2024}, hadron-hadron scattering \cite{Ellis:1980}, heavy quark production \cite{Beenakker:1988}, and single-spin asymmetries \cite{Schlegel:2012, Ringer:2015}.
In processes with massless particles, collinear singularities are present, necessitating a calculation in $d=4-2\eps$ dimensions \cite{tHooft:1972,Bollini:1972}.

The complexity of angular integrals is governed by the number of denominators and masses.
In all applications from the literature mentioned above only integrals with up to two denominators were required.
This was sufficient in these cases, since higher-denominator angular integrals could be reduced to just two-denominator ones by partial fractioning owing to the restricted kinematics with only three observed particles, counting initial and final states.
However, going to processes with a larger number of observed particles leads to a larger number of linearly independent momenta requiring genuine angular integrals with three denominators.

While the two-denominator case is well understood with analytic results in $d$ dimensions \cite{Somogyi:2011,Lyubovitskij:2021} as well as $\eps$-expansions to all orders \cite{Lyubovitskij:2021}, far less is known in the case of more denominators.
In \cite{Somogyi:2011}, the massless three-denominator integral has been calculated to order $\eps^0$ and a general Mellin-Barnes representation for angular integrals was given.
Recently, multi-denominator angular integrals have been studied in the limit of small masses in \cite{Smirnov:2024}.
The $\eps$-expansions for the three-denominator integral with an arbitrary number of masses were given to order $\eps^0$ to leading power in the masses.
Further showcasing the newly increased interest in the subject, the $\eps$-expansion of the single-massive four-denominator angular integral has been studied at as an example for a novel regularization scheme with connections to tropical geometry \cite{Salvatori:2024}.

In this paper we go beyond existing results by giving a comprehensive treatment of three-denominator angular integrals with integer denominator powers and an arbitrary number of masses without an expansion in the latter.
Explicit results will be provided up to order $\eps$.

Following the notation of \cite{Somogyi:2011,Lyubovitskij:2021,Smirnov:2024}, the integral family under consideration is
\begin{align}
I_{j_ 1,j_ 2,j_ 3}^{(m)}(v_{12},v_{13},v_{23},v_{11},v_{22},v_{33};\eps)=\int\frac{\dx\Omega_{d-1}(k)}{\Omega_{d-3}}\frac{1}{(v_ 1\cdot k)^{j_1}(v_ 2\cdot k)^{j_2}(v_ 3\cdot k)^{j_3}} \,,
\label{eq: General angular integral}
\end{align}
with normalized $d$-vectors
\begin{align*}
k&=\change{(1,\vec{k})=}(1,\dots,\sin\theta_1\sin\theta_2\cos\theta_3,\sin\theta_1\cos\theta_2,\cos\theta_1)\,,
\\
v_1&=\change{(1,\vec{v}_1)=}(1,\vec{0}_{d-2},\beta_1)\,,\\
v_2&=\change{(1,\vec{v}_2)=}(1,\vec{0}_{d-3},\beta_2\sin\chi_2^{(1)},\beta_2\cos\chi_2^{(1)})\,,\\
v_3&=\change{(1,\vec{v}_3)=}(1,\vec{0}_{d-4},\beta_3\sin\chi_3^{(1)}\sin\chi_3^{(2)},\beta_3\sin\chi_3^{(1)}\cos\chi_3^{(2)},\beta_3\cos\chi_3^{(1)})\,,
\end{align*}
and normalized angular integration measure
\begin{align}
\dx\Omega\equiv\frac{\dx\Omega_{d-1}(k)}{\Omega_{d-3}}=\frac{\Omega_{d-4}}{\Omega_{d-3}}\,\dx\theta_1\dx\theta_2\dx\theta_3\sin^{d-3}\theta_1\sin^{d-4}\theta_2\sin^{d-5}\theta_3\,,
\label{eq: normalized angular integration measure}
\end{align}
where $\Omega_d=2 \pi^{d/2}/\Gamma(d/2)$.
The normalization is chosen in accordance with \cite{Somogyi:2011} to remove factors of the Euler-Mascheroni constant $\gammaeuler$ from the $\eps$-expansion.
\change{Here and in the following the spatial part of a $d$-vector $\xi$ is indicated by an arrow as $\vec{\xi}$.
The dot product between two $d$-vectors denotes the usual Minkowski scalar product, between spatial vectors it means the Euclidean scalar product, i.e. for normalized $d$-vectors $\xi$ and $\eta$ it is $\xi\cdot\eta=1-\vec{\xi}\cdot\vec{\eta}$.}
The integral depends on the invariants $v_{ij}\equiv v_i\cdot v_j$.
The parametric variables $\beta_i$ and $\chi_j^{(i)}$, where $1\leq i\leq 3$ and $i<j\leq 3$, can be expressed through the invariant variables $v_{ij}$ via
\begin{align}
\beta_i&=\sqrt{1-v_{ii}}\,,\;
\cos\chi_j^{(1)}=\frac{1-v_{1j}}{\beta_1\beta_j}\,,\;
\cos \chi_j^{(2)}=\frac{1-v_{2j}-\beta_2\beta_j\cos\chi_2^{(1)}\cos\chi_j^{(1)}}{\beta_2\beta_j\sin\chi_2^{(1)}\sin\chi_j^{(1)}}\,,
\nonumber\\
\cos \chi_j^{(3)}&=\frac{1-v_{3j}-\beta_3\beta_j\left(\cos\chi_3^{(1)}\cos\chi_j^{(1)}+\sin\chi_3^{(1)}\sin\chi_j^{(1)}\cos\chi_3^{(2)}\cos\chi_j^{(2)}\right)}{\beta_3\beta_j\sin\chi_3^{(1)}\sin\chi_j^{(1)}\sin\chi_3^{(2)}\sin\chi_j^{(2)}}\,.
\end{align}
The denominator powers $j_1$, $j_2$, $j_3$ are assumed to be integers in the following.
The superscript $m=0,1,2,3$ characterizes the number of non-zero masses $v_{11}$, $v_{22}, v_{33}$.

For the calculation of the $\eps$-expansion we use the method of differential equations which is well known for the calculation of loop integrals \cite{Kotikov:1990, Remiddi:1997, Gehrmann:1999, Henn:2013, Henn:2014, Badger:2023}.
Generally, the strategy consists of the following steps:
	\begin{itemize}
		\item[1.] Derive recursion relations from integration-by-parts (IBP) identities to express every $I_{j_1,j_2,j_3}$ in a basis of master integrals.
		\item[2.] Derive a system of differential equations for the master integrals.
		\item[3.] Solve the differential equations order-by-order in $\eps$.
	\end{itemize}
In the particular case at hand, the structure of the angular integrals allows for several simplifications to this general approach.
Most importantly, by studying the behavior under dimensional shift $d\rightarrow d+2$ it is possible to reconstruct the pole and finite part of the $\eps$-expansion solely in terms of known two-denominator integrals.
Simultaneously, the dimensional shift can be used for analytic continuation to obtain finite integral representations -- fit for numerical checks -- for each order in the expansion.
Like IBP relations and differential equations, also dimensional shift relations for angular integrals have their counterpart for loop integrals \cite{Tarasov:1996, Lee:2012}.

\begin{figure}
\centering
\includegraphics[width=0.8\textwidth]{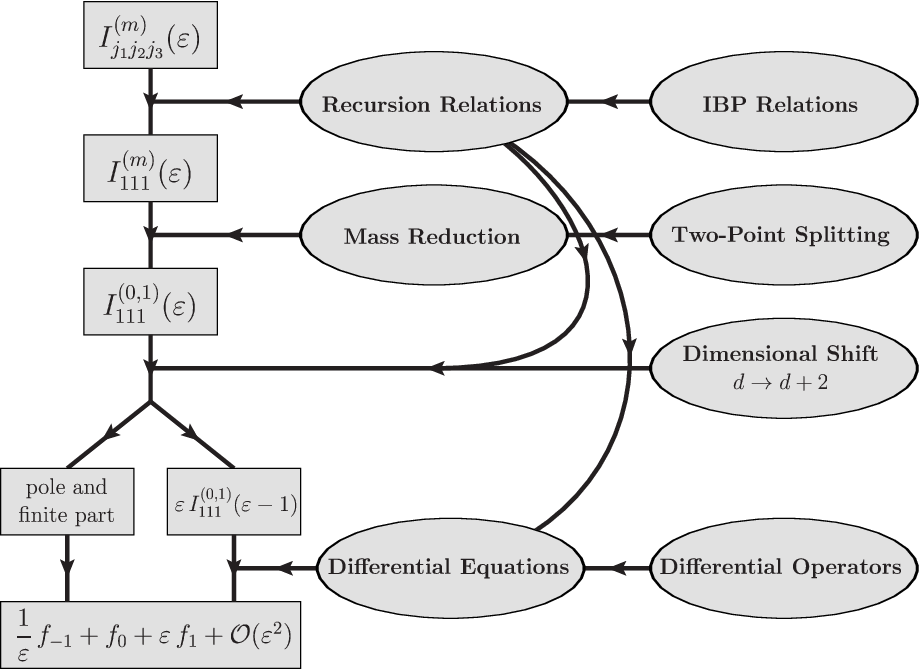}
\caption{This flowchart provides an overview of the calculation of the $\eps$-expansion of the general three-denominator angular integral $I_{j_1j_2 j_3}^{(m)}$. 
In a first step, recursion relations (see app.\,\ref{app: Recursion relations}) derived from IBP relations (see sec.\,\ref{sec: IBP relations}) are used for a reduction to the master integral $I_{1,1,1}^{(m)}$. 
In a second step, the double- and triple-massive integrals are expressed in terms of massless and single-massive ones through mass reduction formulae derived from the two-point splitting lemma (see sec.\,\ref{sec: Reduction of masses}).
In a third step, a combination of a dimensional-shift identity, relating integrals in $d$ and $d+2$ dimensions, with the recursion relations allows for the determination of pole and finite part in terms of known two-denominator integrals (see sec.\,\ref{sec: Dimensional shift identity}). 
In a final step, the order $\eps$ contribution is calculated by applying the method of differential equations -- requiring suitable differential operators for angular integrals (see sec.\,\ref{sec: Differential operators for angular integrals}) and again making use of the recursion relations -- to the massless and single-massive master integral in $d=6-2\eps$ dimensions (see sec.\,\ref{sec: Differential equations}  and \ref{sec: Integration for master integrals}).
Results are found in sec.\,\ref{sec: Results}.
Image created with JaxoDraw \cite{Binosi:2003}.}
\label{fig: Flowchart of calculation}
\end{figure}

To facilitate keeping an overview of what follows, figure \ref{fig: Flowchart of calculation} provides a graphical overview of the steps taken to extract the $\eps$-expansion of the general three-denominator angular integral with $m$ masses $I_{j_1,j_2,j_3}^{(m)}$.

At this point we like to briefly comment that the most common way for the analytic integration of phase space integrals is the method of reversed unitarity \cite{Anastasiou:2002,Anastasiou:2003,Anastasiou:2004}. 
This directly translates phase-space integrals to loop integrals,  opening up the toolkit developed for these including IBP reduction and the derivation of differential equations, e.g. \texttt{FIRE} \cite{Smirnov:2019}, \texttt{KIRA} \cite{Klappert:2020} , and \texttt{LiteRed} \cite{Lee:2013}.
In our calculation we will however perform the procedure of IBP reduction and derivation of differential equations directly on the level of angular integrals without a mapping to loop integrals, showcasing that the introduction of auxiliary cut propagators is not needed for the specific phase-space integrals under consideration.
Since the integral family has a sufficiently simple structure it is feasible to proceed without the aforementioned powerful machinery.
For an example of the treatment of very similar integrals with reversed unitarity technology we refer to \cite{Liu:2024}, for an example of a class of phase-space integrals which inherently require extensions of reversed unitarity to perform the IBP reduction see \cite{Baranowski:2022}.

The remainder of the paper is organized as follows.
In section \ref{sec: IBP relations}, we derive IBP relations to reduce $I_{j_1,j_2,j_3}$ to master integrals.
Section \ref{sec: Reduction of masses} recalls the two-point splitting lemma, which allows us to express double- and triple-massive integrals in terms of the single-massive integral.
The dimensional-shift identity introduced in section \ref{sec: Dimensional shift identity} allows us to express the pole and finite part of the three-denominator integral solely in terms of known two-denominator integrals.
In section \ref{sec: Differential operators for angular integrals}, we determine differential operators, which we subsequently use to derive differential equations for the master integrals in section \ref{sec: Differential equations}.
The integration of these differential equations is discussed in section \ref{sec: Integration for master integrals}.
Results for the massless, single-massive, double-massive, and triple-massive master integrals $I^{(0,1,2,3)}_{1,1,1}$ are given in section \ref{sec: Results}.
The generalization towards angular integrals with more than three denominators is briefly discussed in section \ref{sec: Beyond three denominators}, before we conclude in section \ref{sec: Conclusion}.
Appendix \ref{app: Scalar products for mass reduction formula} lists the scalar products used for the mass reduction formula, while appendices \ref{app: IBP relations} and \ref{app: Recursion relations} give the full form of the IBP relations and the recursion relations obtained from the latter.
In appendix \ref{app: Proof of the general dimensional shift identity}, a proof of the general dimensional-shift identity for angular integrals is provided.
Appendix \ref{app: Expansion of dimensionally shifted two-denominator integrals} lists the expansions for the dimensionally-shifted two-denominator integrals that are used as input for the calculation of the three-denominator master integrals.
The explicit calculation of the boundary value for the massless master integral is given in appendix \ref{app: Calculation of the boundary value for the massless master integral}.
Integrals useful for the calculation of the master integrals are found in appendix \ref{app: Useful integrals for the calculation of the master integrals}, and appendix \ref{app: Clausen function} recalls the Clausen function appearing in the order $\eps$ of the three-denominator angular integrals.
	The ancillary \texttt{Mathematica} file provides the function \verb!AngularIntegral[{j1_,j2_,j3_},{v12_,v13_,v23_,v11_,v22_,v33_}]! which evaluates the three-denominator angular integral in $d=4-2\eps$ dimensions up to order $\eps$.

\section{IBP relations}
\label{sec: IBP relations}
Since they are the foundation for everything that follows, we \change{start} by deriving IBP relations to reduce $I_{j_1,j_2,j_3}$ to master integrals.
IBP relations for angular integrals are derived in a similar fashion to loop integrals.
The only difference is that permissible vectors in the IBP relation have to be tangential to the spherical integration surface.
Concretely, the IBPs read
\begin{align}
0=\int \mathrm{d}\Omega \frac{\partial}{\partial\vec{k}}\cdot\vec{\xi}\
\frac{1}{(v_ 1\cdot k)^{j_ 1}(v_ 2\cdot k)^{j_ 2}(v_ 3\cdot k)^{j_ 3}} \,,
\label{eq: general IBP}
\end{align}
where  the \change{spatial} vector $\vec{\xi}$ must satisfy
$\vec{\xi}\cdot\vec{k}\overset{!}{=}0$.

To construct possible vectors $\vec{\xi}$ we can use the $\vec{v}_i$ as well as $\vec{k}$.
From the orthogonality condition, we find  the three vectors ($i=1,2,3$; note that $k$ is not an index)
\begin{align}
\vec{\xi}_{ik}=(\vec{v}_i\cdot \vec{k})\vec{k}-\vec{k}^2 \vec{v}_i
\label{eq: Def xi_ik}
\end{align}
and the additional three vectors ($i<j$)
\begin{align}
\vec{\xi}_{ij}=(\vec{v}_i\cdot \vec{k})\vec{v}_j-(\vec{v}_j\cdot \
\vec{k})\vec{v}_i\,.
\label{eq: Def xi_ij}
\end{align}
For these vectors, it is
\begin{align}
\frac{\partial\cdot \vec{\xi}_{ik}}{\partial\vec{k}}=(d-2)
\vec{v}_i\cdot \vec{k}\quad\text{respectively}\quad\frac{\partial\cdot\vec{\xi}_{ij}}{\partial\vec{k}}=0
\label{eq: Differentiate xi}
\end{align}
and furthermore, recalling $v_i\cdot k=1-\vec{v}_i\cdot\vec{k}$,
\begin{align}
\frac{\partial}{\partial\vec{k}}\frac{1}{(v_i\cdot k)^{j_i}}=\frac{j_i \vec{v}_i}{(v_i\cdot k)^{j_i+1}}\,.
\label{eq: Differentiate propagator}
\end{align}
Applying the product rule, eq.\,\eqref{eq: general IBP} becomes
	\begin{align}
		0=\int \dx\Omega&\left[\left(\frac{\partial\cdot\vec{\xi}}{\partial\vec{k}}\right)\frac{1}{(v_ 1\cdot k)^{j_ 1}(v_2\cdot k)^{j_ 2}(v_3\cdot k)^{j_3}}
		%\right.
		%\nonumber\\
		%&\left.
		+\vec{\xi}\cdot\frac{\partial}{\partial\vec{k}}\frac{1}{(v_ 1\cdot k)^{j_ 1}(v_2\cdot k)^{j_ 2}(v_3\cdot k)^{j_3}}\right] .
		\label{eq: IBP relations product rule}
	\end{align}
Using eqs.~\eqref{eq: Differentiate xi} and \eqref{eq: Differentiate propagator} on eq.~\eqref{eq: IBP relations product rule} and expressing all scalar products in terms of (inverse) propagators leads to an IBP relation for each of the six $\vec{\xi}$.
Those relate $I_{j_1,j_2,j_3}$ for different indices.
We list them explicitly in appendix \ref{app: IBP relations}.

For general $j_{1,2,3}$ these IBP relations can be combined into identities that only either raise or lower the sum of indices.
This allows to bring them in the form of recursion relations that systematically reduce all integrals $I_{j_1,j_2,j_3}$ into the master integrals $I_{1,1,1}$, $I_{1,1,0}$, $I_{1,0,0}$, $I_{0,0,0}$ (+ permutation of indices), bypassing the need for Laporta's algorithm \cite{Laporta:2000}.
The explicit recursion relations are given in appendix \ref{app: Recursion relations}.

To calculate the master integrals, we use the method of differential equations.
For this, we write the master integrals in the form of eq.\,\eqref{eq: General angular integral} and differentiate both sides with respect to the kinematic invariants $v_{ij}$, which will lead to new angular integrals on the right-hand side -- details of the differentiation are discussed in section \ref{sec: Differential operators for angular integrals}.
By reducing the right-hand side to master integrals via the recursion relations, we obtain a system of differential equations for the master integrals.
For a pedagogical introduction to the differential equation approach in the context of loop integrals see \cite{Henn:2014, Badger:2023}.

Since the $\eps$-expansion of two-denominator angular integrals is known \cite{Beenakker:1988, Somogyi:2011,Lyubovitskij:2021}, the only unknown master integral in our case is $I_{1,1,1}$.
By explicitly inserting the known integrals into the coupled system of differential equations, it is reduced to one closed differential equation for $I_{1,1,1}$.

To further simplify matters we use the following tricks:
\begin{itemize}
\item[(a)] By two-mass splitting we can always reduce the triple-massive integral to single-massive integrals.
We will be discussing this next in section \ref{sec: Reduction of masses}. 
This simplifies the differential equation by reducing the number of variables. The number of master integrals is further reduced since the massless one-denominator integral \change{$I^{(0)}_{1,0,0}$} is reducible to \change{$I^{(0)}_{0,0,0}$}.
\item[(b)] By using a dimensional-shift identity $d\rightarrow d+2$ we obtain the pole and -- somewhat surprisingly -- also the finite part of the three-denominator integral ``for free", i.e. in terms of known integrals with fewer denominators.
We will look at this in section \ref{sec: Dimensional shift identity}.
\item[(c)] By symmetry it suffices to look at a single derivative, say ${\partial}/{\partial v_{12}}$, for the massless case.
From there we only need a single boundary value for which we choose the symmetric point $v_{12}=v_{13}=v_{23}=1$.
To construct the single-massive integral from the differential equation with respect to ${\partial}/{\partial v_{11}}$ we can use the previously calculated massless integral as a boundary condition.
This will be done in sections \ref{sec: Differential equations} and \ref{sec: Integration for master integrals}.
\end{itemize}

\section{Mass reduction via two-point splitting}
\label{sec: Reduction of masses}
It is a truth universally acknowledged, that a single-massive angular integral is sufficient to calculate multi-mass angular integrals.
This is possible due to the two-point splitting lemma \cite{Lyubovitskij:2021, Wunder:2024}, which  states that for any two vectors $v_1$ and $v_2$, we can choose any scalar $\lambda_{(12)}$ and construct the linear combination
${
v_{(12)}\equiv(1-\lambda_{(12)})\,v_1+\lambda_{(12)}\,v_2
}$
to obtain 
	\begin{align}
		\frac{1}{v_1\cdot k\,\,v_2\cdot k}=\frac{\lambda_{(12)}}{v_1\cdot k\,\,v_{(12)}\cdot k}+\frac{1-\lambda_{(12)}}{v_2\cdot k\,\,v_{(12)}\cdot k}\,.
		\label{eq: Two point splitting lemma}
	\end{align}
If we consider two massive vectors $v_1$ and $v_2$ and choose $\lambda_{(12)}$ such that $v_{(12)}$ is massless, i.e. $v_{(12)(12)}=0$, eq.\eqref{eq: Two point splitting lemma} splits the product of two massive denominators into the sum of two single-massive denominator products.
Iterative use of this identity on pairs of massive denominators allows for a reduction to the single-massive case.
Hence angular integrals with an arbitrary number of masses can always be expressed through at most single-massive integrals.\,\footnote{A quite similar idea of splitting loop integrals into integrals with fewer masses was already put forward a long time ago by 't Hooft and Veltman \cite{tHooft:1978}.}

Solving the masslessness condition $v_{(12)(12)}\overset{!}{=}0$ for $\lambda_{(12)}$ yields the solutions
\begin{align}
\lambda_{(12)}^\pm=\frac{v_{12}-v_{11}\pm\sqrt{v_{12}^2-v_{11}v_{22}}}{2v_{12}-v_{11}-v_{22}}\,.
\end{align}
Both of these solutions can be used for a reduction of the number of masses.
To make the reduction well-behaved in the massless limit we can bring eq.\eqref{eq: Two point splitting lemma} into symmetric form by introducing the additional vector $v_{(21)}\equiv\lambda_{(21)}\, v_1+(1-\lambda_{(21)})\,v_2$.
Employing the two-point splitting lemma \,\eqref{eq: Two point splitting lemma} first on $1/(v_1\cdot k\,v_2\cdot k)$ inserting $v_{(12)}$ and subsequently on $1/(v_2\cdot k\,v_{(12)}\cdot k)$ inserting $v_{(21)}$, we obtain the splitting
\begin{align}
\frac{1}{v_1\cdot k\,\,v_2\cdot k}=&\frac{\lambda_{(12)}}{v_1\cdot k\,\,v_{(12)}\cdot k}+\frac{\lambda_{(21)}}{v_2\cdot k\,\,v_{(21)}\cdot k}
+\frac{1-\lambda_{(12)}-\lambda_{(21)}}{v_{(12)}\cdot k\,\,v_{(21)}\cdot k}\,.
\label{eq: Two point splitting lemma double massive two denominator integral}
\end{align}
To make $v_{(12)}$ and $v_{(21)}$ massless as well as coinciding with $v_1$ respectively $v_2$ in the respective massless limit, we choose
\begin{align}
\lambda_{(12)}=\frac{v_{12}-v_{11}-\sqrt{X_{12}}}{2v_{12}-v_{11}-v_{22}}\,,\quad
\lambda_{(21)}=\frac{v_{12}-v_{22}-\sqrt{X_{12}}}{2v_{12}-v_{11}-v_{22}}\,,
\end{align}
with $X_{12}\equiv 
%-\det\!\left(\begin{tiny}
%\begin{array}{cc}
%v_{11} &v_{12}\\ v_{12} &v_{22}
%\end{array}\end{tiny}\right)=
v_{12}^2-v_{11}v_{22}$.
Note that there is the geometrical interpretation $X_{12}=4\,\vol(\vec{v}_1,\vec{v}_2)^2+(\vec{v}_1-\vec{v}_2)^2$, where $\vol(\vec{v}_1,\vec{v}_2)$ denotes the area of the triangle spanned by $\vec{v}_1$ and $\vec{v}_2$.
Hence, for real vectors $\vec{v}_{1,2}$, $X_{12}$ is non-negative and vanishes if and only if $\vec{v}_1=\vec{v}_2$.
The scalar products of the auxiliary vectors are ${v_{(12)(12)}=v_{(21)(21)}=0}$, ${v_{1(12)}=-\lambda_{(12)}\sqrt{X_{12}}}$, ${v_{2(21)}=-\lambda_{(21)}\sqrt{X_{12}}}$, and ${v_{(12)(21)}=2X_{12}/(2v_{12}-v_{11}-v_{22})}$.
Note that $v_{1(12)}$ respectively $v_{2(21)}$ vanish in the massless limits $v_{11} \rightarrow 0$ respectively $v_{22} \rightarrow 0$ .
Using these scalar products, a compact form of the two-mass splitting is
\begin{align}
\frac{1}{v_1\cdot k\,\,v_2\cdot k}=&\frac{1}{\sqrt{X_{12}}}\left[\frac{v_{(12)(21)}}{v_{(12)}\cdot k\,\, v_{(21)}\cdot k}-\frac{v_{1(12)}}{v_1\cdot k\,\, v_{(12)}\cdot k}-\frac{v_{2(21)}}{v_2\cdot k\,\, v_{(21)}\cdot k}\right].
\label{eq: Two point splitting lemma double massive two denominator integral compact}
\end{align}

The reduction formula for the double-massive master integral immediately following from this is
\begin{align}
&I_{1,1,1}^{(2)}(v_{12},v_{13},v_{23},v_{11},v_{22})=\frac{1}{\sqrt{X_{12}}}\left[ v_{(12)(21)}
I_{1,1,1}^{(0)}\left(v_{3(12)},v_{3(21)},v_{(12)(21)}\right)
\right.  \nonumber 
\\
&\qquad\qquad\left.
-v_{1(12)}I_{1,1,1}^{(1)}\left(v_{13},v_{1(12)},v_{3(12)},v_{11}\right)
-v_{2(21)}I_{1,1,1}^{(1)}\left(v_{23},v_{2(21)},v_{3(21)},v_{22}\right)
\right]
\label{eq: reduction of two-mass integral}
\end{align}
In case $v_3$ is massive as well, eq.\eqref{eq: Two point splitting lemma double massive two denominator integral} is applied twice to obtain the following reduction formula for the triple-massive master integral
\begin{align}
I_{1,1,1}^{(3)}&(v_{12},v_{13},v_{23},v_{11},v_{22},v_{33})=
\frac{1}{\sqrt{X_{12}}}
\Big\lbrace
v_{(12)(21)}
I_{1,1,1}^{(1)}\left(v_{3(12)},v_{3(21)},v_{(12)(21)},v_{33}\right)
\nonumber\\
&
-\frac{v_{1(12)}}{\sqrt{X_{13}}}
\Big[
 v_{(13)(31)}
I_{1,1,1}^{(0)}\left(v_{(12)(13)},v_{(12)(31)},v_{(13)(31)}\right)
\nonumber\\
&\qquad
-v_{1(13)}I_{1,1,1}^{(1)}\left(v_{1(12)},v_{1(13)},v_{(12)(13)},v_{11}\right)
-v_{3(31)}I_{1,1,1}^{(1)}\left(v_{3(12)},v_{3(31)},v_{(12)(31)},v_{33}\right)
\Big]
\nonumber\\
&
-\frac{v_{2(21)}}{\sqrt{X_{23}}}
\Big[
 v_{(23)(32)}
I_{1,1,1}^{(0)}\left(v_{(21)(23)},v_{(21)(32)},v_{(23)(32)}\right)
\nonumber\\
&\qquad-v_{2(23)}I_{1,1,1}^{(1)}\left(v_{2(21)},v_{2(23)},v_{(21)(23)},v_{22}\right)
-v_{3(32)}I_{1,1,1}^{(1)}\left(v_{3(21)},v_{3(32)},v_{(21)(32)},v_{33}\right)
\Big]
\Big\rbrace\,,
\label{eq: reduction of three-mass integral}
\end{align}
where the explicit expressions for the scalar products that appear are listed in appendix \ref{app: Scalar products for mass reduction formula}.
In principle one could derive reduction formulae for any integer indices $j_i$, however in light of the recursion relations derived in the previous section and given in appendix \ref{app: Recursion relations} this is not required.

We note that we can present the mass reduction formulae in a much cleaner form by scaling the three-denominator integrals by their Euclidean Gram determinant
\begin{align}
Y_{ijk}=\det\!\left(\begin{array}{ccc}
1-v_{ii} & 1-v_{ij} & 1-v_{ik} \\
1-v_{ij} & 1-v_{jj} & 1-v_{jk} \\
1-v_{ik} & 1-v_{jk} & 1-v_{kk} 
\end{array}\right).
\label{eq: Euclidean Gram Determinant}
\end{align}
Defining
\begin{align}
\tilde{I}^{(m)}_{1,1,1}(v_{ij},v_{ik},v_{jk},v_{ii},v_{jj},v_{kk})\equiv\sqrt{Y_{ijk}} \,I^{(m)}_{1,1,1}(v_{ij},v_{ik},v_{jk},v_{ii},v_{jj},v_{kk}) \,,
\end{align}
the two-mass reduction formula \eqref{eq: reduction of two-mass integral} takes the form
\begin{align}
\tilde{I}_{1,1,1}^{(2)}(v_{12},v_{13},v_{23},v_{11},v_{22})=&
\tilde{I}_{1,1,1}^{(0)}\left(v_{3(12)},v_{3(21)},v_{(12)(21)}\right)
-\tilde{I}_{1,1,1}^{(1)}\left(v_{13},v_{1(12)},v_{3(12)},v_{11}\right)
\nonumber\\
&-\tilde{I}_{1,1,1}^{(1)}\left(v_{23},v_{2(21)},v_{3(21)},v_{22}\right),
\label{eq: reduction of two-mass integral simplified}
\end{align}
which remarkably is free of explicit kinematic prefactors.
This generalizes to the three-mass reduction \eqref{eq: reduction of three-mass integral} which for the scaled angular integrals becomes
\begin{align}
\tilde{I}_{1,1,1}^{(3)}&(v_{12},v_{13},v_{23},v_{11},v_{22},v_{33})=
\tilde{I}_{1,1,1}^{(1)}\left(v_{3(12)},v_{3(21)},v_{(12)(21)},v_{33}\right)
\nonumber\\
&-\tilde{I}_{1,1,1}^{(0)}\left(v_{(12)(13)},v_{(12)(31)},v_{(13)(31)}\right)
+\tilde{I}_{1,1,1}^{(1)}\left(v_{1(12)},v_{1(13)},v_{(12)(13)},v_{11}\right)
\nonumber\\
&
+\tilde{I}_{1,1,1}^{(1)}\left(v_{3(12)},v_{3(31)},v_{(12)(31)},v_{33}\right)
-\tilde{I}_{1,1,1}^{(0)}\left(v_{(21)(23)},v_{(21)(32)},v_{(23)(32)}\right)
\nonumber\\
&
+\tilde{I}_{1,1,1}^{(1)}\left(v_{2(21)},v_{2(23)},v_{(21)(23)},v_{22}\right)
+\tilde{I}_{1,1,1}^{(1)}\left(v_{3(21)},v_{3(32)},v_{(21)(32)},v_{33}\right).
\label{eq: reduction of three-mass integral simplified}
\end{align}

\section{Dimensional-shift identity}
\label{sec: Dimensional shift identity}
The $n$-denominator angular integral satisfies the general dimensional-shift identity
\begin{align}
I_{j_ 1,...,j_n}(\eps)=&\frac{1}{1-2\eps}\left[\sum_{i=1}^n j_i I_{j_1,...,j_i+1,...,j_n}(\eps-1)
%\right.
%\nonumber\\
%&\left.
+\,\left(3-\sum_{i=1}^n j_i-2\eps\right)I_{j_1,...,j_n}(\eps-1)\right],
\label{eq: General dimensional shift identity}
\end{align}
which expresses the integral in $d=4-2\eps$ dimensions on the left side as a sum over integrals in $d+2=6-2\eps$ dimensions on the right.
A proof of eq.\,\eqref{eq: General dimensional shift identity} is given in appendix \ref{app: Proof of the general dimensional shift identity}.

In combination with IBP reduction, this identity can be used to analytically continue the angular integral from a convergent dimension to a divergent one.
Note that for the three-denominator angular integral already the angular measure as defined in eq.\,\eqref{eq: normalized angular integration measure} is divergent in $d=4$ dimensions due to the $\sin^{-1}\theta_3$ factor.
With the above formula we can however express the integral in $d=4$ dimensions in terms of integrals in $d=6$ dimensions where the measure is finite because it contains an additional $\sin^2\theta_3$ factor.

Also the master integrals with $j_i=1$ and $v_{ii}=0$, which are collinearly divergent in $d=4$ dimensions, converge for sufficiently large dimension, i.e. small $\eps$. 
In particular they are finite in $d=6$ dimensions.
Hence applying eq.\eqref{eq: General dimensional shift identity} to $I_{1,1,1}(\eps)$ and subsequently reducing the integrals in six dimensions on the right to master integrals, results in all $1/\eps$ poles being captured in prefactors of the master integrals.

As it turns out the identity connecting $I_{1,1,1}(\eps)$ and $I_{1,1,1}(\eps-1)$ simplifies further by shifting all integrals with two or less denominators back to $d=4-2\eps$ dimensions.
The necessary identities follow from eq.\eqref{eq: General dimensional shift identity} by subsequently choosing zero, one, and two denominators, performing an IBP reduction and solving for the $d=6-2\eps$ dimensional integrals.
Going through the outlined steps, we find the remarkably simple relation
\begin{align}
I_{1,1,1}(\eps)=\frac{1}{X_{123}}\left[X_{12\bar{3}}I_{1,1,0}(\eps)+X_{1\bar{2}3}\,I_{1,0,1}(\eps)+X_{\bar{1}23}\,I_{0,1,1}(\eps)+\frac{2\eps}{1-2\eps}\,Y_{123}\,I_{1,1,1}(\eps-1)\right].
\label{eq: I111 dimensional shift representation}
\end{align}
The kinematic factor $X_{123}$ in this relation is given by the (Minkowski-)Gram determinant
\begin{align}
X_{123}\equiv&\det\!\left(
\begin{array}{ccc}
v_{11} & v_{12} & v_{13} \\
v_{12} & v_{22} & v_{23} \\
v_{13} & v_{23} & v_{33} 
\end{array}
\right)
=2 v_{12} v_{13} v_{23}-v_{11} v_{23}^2-v_{13}^2 v_{22} -v_{33} v_{12}^2+v_{11} v_{22} v_{33}\,.
\end{align}
The factors $X_{12\bar{3}}$, $X_{1\bar{2}3}$, and $X_{\bar{1}23}$ multiplying the two-denominator integrals are obtained by interchanging the barred column in the Gram matrix with a column of ones\footnote{Reminiscent of Cramer's rule for the solution of linear equations.} , i.e.
\begin{align}
X_{12\bar{3}}\equiv\det\!\left(
\begin{array}{ccc}
v_{11} & v_{12} & 1 \\
v_{12} & v_{22} & 1 \\
v_{13} & v_{23} & 1
\end{array}
\right)=v_{12} (v_{13}+ v_{23}-v_{12})-v_{11} v_{23}-v_{22} v_{13}+v_{11} v_{22},
\end{align}
for $X_{1\bar{2}3}$ and $X_{\bar{1}23}$ indices are interchanged cyclically.
The factor $Y_{123}$ is given by the Euclidean Gram determinant already encountered in eq.\,\eqref{eq: Euclidean Gram Determinant},
\begin{align}
Y_{123}
%\equiv
%&\det\!\left(
%\begin{array}{ccc}
%\vec{v}_1^2 & \vec{v}_1\cdot\vec{v}_2 & \vec{v}_1\cdot\vec{v}_3 \\
%\vec{v_{1}}\cdot\vec{v}_2 & \vec{v}_2^2 & \vec{v}_2\cdot\vec{v}_3\\
%\vec{v}_{1}\cdot\vec{v}_3 & \vec{v}_2\cdot\vec{v}_3 & \vec{v}_3^2
%\end{array}
%\right)
%\nonumber\\
=&2 v_{12}v_{13}+2v_{12}v_{23}+2v_{13}v_{23}-v_{12}^2-v_{13}^2-v_{23}^2-2v_{12}v_{13}v_{23}
-v_{11}v_{23}(2-v_{23})
\nonumber
\\
&
-v_{22}v_{13}(2-v_{13})
-v_{33}v_{12}(2-v_{12})
+v_{11}v_{22}+v_{11}v_{33}+v_{22}v_{33}-v_{11}v_{22}v_{33}\,.
\label{eq: Euclidean Gram Determinant Y123}
\end{align}
Note, that geometrically
\begin{align}
Y_{123}=\beta_1^2\beta_2^2\beta_3^2\sin^2\chi_2^{(1)}\sin^2\chi_3^{(1)}\sin^2\chi_3^{(2)}=36\,\vol^2(\vec{v}_1,\vec{v}_2,\vec{v}_3)
\label{eq: Y123 geometrically}
\end{align}
where $\vol(\vec{v}_1,\vec{v}_2,\vec{v}_3)$ denotes the Euclidean volume of the tetrahedron spanned by the vectors $\vec{v}_{1,2,3}$.
We note that for real valued vectors $\vec{v}_i$ the quantity $Y_{123}$ appearing in the denominator vanishes if and only if the vectors are confined to a plane, i.e. they are linearly dependent.
In this particular case eq.~\eqref{eq: I111 dimensional shift representation} becomes a reduction identity of the three-denominator integral to a sum of two-denominator integrals --
remarkably the very same, one would also find by partial fractioning of the integrand.

Using the formula \eqref{eq: I111 dimensional shift representation} has two benefits.
Firstly for the purpose of analytical calculation, we gain the pole part and the $\eps^0$ term ``for free" since they are contained entirely in terms of known two-denominator integrals.
Since $I_{1,1,1}(\eps-1)$ is finite for $\eps=0$ and is multiplied by an explicit $\eps$ in eq.~\eqref{eq: I111 dimensional shift representation}, it only starts contributing at order $\eps$.
Hence, for results to order $\eps$ it suffices to calculate the leading term of $I_{1,1,1}(\eps-1)$ by differential equations.

Secondly, from a numerical perspective, by expanding the integration measure of $I_{1,1,1}(\eps-1)$ in a series in $\eps$ we obtain a finite integral representation for every order in $\eps$.
In conjunction with the known expansion of the two-denominator integrals this can be used for numeric checks.

\section{Differential operators for angular integrals}
\label{sec: Differential operators for angular integrals}
To derive differential equations with respect to the scalar products $v_{ij}$, we need to differentiate 
	\begin{align}
		\int \mathrm{d}\Omega\,\frac{1}{(v_ 1\cdot k)^{j_1}(v_2\cdot k)^{j_2}(v_3\cdot k)^{j_3}} \,,
	\end{align}
which however depends on the explicit vectors $\vec{v}_i$. Thus, the first step is to derive appropriate differential operators. 
Due to symmetry, it is sufficient to consider the derivatives with respect to $v_{12}=1-\vec{v}_1\cdot \vec{v}_2$ respectively the mass $v_{11}=1-\vec{v}_1\cdot \vec{v}_1$.
For this we make the ans\"atze
	\begin{align}
		\frac{\partial}{\partial v_{12}}\overset{!}{=}(a_1 \vec{v}_1+a_2 \vec{v}_2+a_3\vec{v}_3)\cdot\frac{\partial}{\partial \vec{v_1}}\,\quad
		%\\
		\text{respectively}\quad
		\frac{\partial}{\partial v_{11}}\overset{!}{=}(b_1 \vec{v}_1+b_2 \vec{v}_2+b_3\vec{v}_3)\cdot\frac{\partial}{\partial \vec{v_1}}\,.
	\end{align}
The coefficients $a_{1,2,3}$ and $b_{1,2,3}$ are fixed by the conditions
	\begin{align}
		&\frac{\partial v_{11}}{\partial v_{12}}=0\,,\quad\frac{\partial v_{12}}{\partial v_{12}}=1\,,\quad\frac{\partial v_{13}}{\partial v_{12}}=0\quad
		%\\
		\text{respectively}\quad
		\frac{\partial v_{11}}{\partial v_{11}}=1\,,\quad\frac{\partial v_{12}}{\partial v_{11}}=0\,,\quad\frac{\partial v_{13}}{\partial v_{11}}=0\,.
	\end{align}
Derivatives of the other invariants $v_{22}$, $v_{23}$, $v_{33}$ are trivially $0$ due to the chosen ansatz.
For $\partial/\partial v_{12}$, in the massless case, where $v_{11}=v_{22}=v_{33}=0$, we find
\begin{align}
\frac{\partial}{\partial v_{12}}=\frac{1}{Y_{123}^{(0)}}&\Big[(v_{13}+v_{23}-v_{12}-v_{13}v_{23})\vec{v}_1
\nonumber\\
&
-(2-v_{13})v_ {13}\vec{v}_2
+(v_{12}+v_{13}-v_{23}-v_{12}v_{13})\vec{v}_3\Big]\cdot\frac{\partial}{\partial\vec{v}_1},
\end{align}
where in the massless case the denominator is given by
\begin{align}
Y_{123}^{(0)}=&-v_{12}^2-v_ {13}^2-v_{23}^2+2v_{12}v_{13}+2v_{12}v_{23}
%\nonumber\\
%&
+2v_{13}v_{23}-2 v_{12}v_{13}v_{23},
\label{eq: massless X}
\end{align}
which is the massless Euclidean Gram determinant from eq.\,\eqref{eq: Euclidean Gram Determinant Y123}.
Of course the operator can be generalized to the case of non-zero masses, however this is not needed in the following.
The differential operators $\partial/\partial v_{13}$ and $\partial/\partial v_{23}$ follow by interchanging indices.

For the differential operator $\partial/\partial v_{11}$
we find in the single-massive case $v_{22}=v_{33}=0$
\begin{align}
\frac{\partial}{\partial v_{11}}=
\frac{1}{2 Y_{123}^{(1)}}&\Big[
-v_{23}(2-v_{23})\vec{v}_1
%\nonumber\\
%&
+(v_{13}+v_{23}-v_{12}-v_{13}v_{23})\vec{v}_2
\nonumber\\
&
+(v_{12}+v_{23}-v_{13}-v_{12}v_{13})\vec{v}_3
\Big]\cdot\frac{\partial}{\partial\vec{v}_1},
\label{eq: single-massive X}
\end{align}
where the denominator is given by
\begin{align}
Y_{123}^{(1)}=Y_{123}^{(0)}-v_{11}v_{23}(2-v_{23})\,.
\end{align}
Again we recognize the Euclidean Gram determinant from eq.~\eqref{eq: Euclidean Gram Determinant Y123}, this time for one non-zero mass.
\section{Differential equations}
\label{sec: Differential equations}
Having constructed suitable differential operators we are now equipped to derive differential equations for the three-denominator angular integral.
The only master integral with genuinely three denominators is $I_{1,1,1}$.
All other master integrals have at most two denominators and are known to all orders in $\eps$.
To construct the full solution for an arbitrary number of masses, we have to look at two cases.
First we consider the massless integral.
It is symmetric in its variables $v_{12}, v_{13}, v_{23}$, hence it suffices to only consider the differential equation $\partial/ \partial v_{12}$ and construct a symmetric solution.
As a boundary value we will use the integral at the symmetric point $v_{12}=v_{13}=v_{23}=1$.
The  second case to consider is the single-massive integral with $v_{11}\neq 0$.
Here, we look at the differential equation with respect to $\partial/\partial v_{11}$ and use the massless integral as a boundary condition.
As discussed in section \ref{sec: Dimensional shift identity} it is beneficial to consider the differential equations for integrals in $d=6-2\eps$ dimensions.
This way, the leading-order solution in $\eps$ of the differential equation, which is of order $\eps^0$, will provide a contribution to the order $\eps$ of the master integral in $d=4-2\eps$ dimensions.

By applying the differential operators to the massless integral, we obtain
\begin{align}
\frac{\partial I^{(0)}_{1,1,1}}{\partial v_{12}}=\frac{1}{Y_{123}^{(0)}}&\Big[
v_{13}(v_{13}-v_{12}-v_{23})I^{(0)}_{2,1,1}
%\nonumber\\
%&
+(v_{12}-v_{13}-v_{23}+v_{13}v_{23})I^{(0)}_{1,1,1}
%\nonumber\\
%&
+v_{13}(2-v_{13})I^{(0)}_{2,0,1}
\nonumber\\
&
+(v_{23}-v_{12}-v_{13}+v_{12}v_{13})I^{(0)}_{2,1,0}
\Big]
\end{align}
and for the single-massive integral
\begin{align}
\frac{\partial I^{(1)}_{1,1,1}}{\partial v_{11}}=\frac{1}{2Y_{123}^{(1)}}&\Big[
v_{23}(v_{23}-v_{12}-v_{13})I^{(1)}_{2,1,1}
%\nonumber\\
%&
+v_{23}(2-v_{23})I^{(1)}_{1,1,1}
%\nonumber\\
%&
+(v_{13}-v_{12}-v_{23}+v_{12}v_{23})I^{(1)}_{2,1,0}
\nonumber\\
&
+(v_{12}-v_{13}-v_{23}+v_{13}v_{23})I^{(1)}_{2,0,1}
\Big].
\end{align}
From here, we can use the recursion relations from appendix \ref{app: Recursion relations} to reduce the right-hand side of both equations to master integrals.
In both cases, this results in a closed differential equation for $I_{1,1,1}$, where all other terms are expressed in terms of known lower-denominator integrals.

The reduction to master integrals of the massless equation in $d$ dimensions leads to
\begin{align}
\frac{\partial I^{(0)}_{1,1,1}}{\partial v_{12}}&=
\left[\frac{d-6}{2v_{12}}-\frac{d-5}{Y_{123}^{(0)}}\,(v_{13}+v_{23}-v_{23}-v_{13}v_{23})\right]I^{(0)}_{1,1,1}
\nonumber\\
&+\frac{1}{2v_{12}Y_{123}^{(0)}}
\Big[(d-5)v_{12}(4-v_{12}-v_{13}-v_{23})I^{(0)}_{1,1,0}
\nonumber\\
&+(d-5)(2-v_{13})(v_{23}-v_{12}-v_{13})I^{(0)}_{1,0,1}
+(d-5)(2-v_{23})(v_{13}-v_{12}-v_{23})I^{(0)}_{0,1,1}
\Big].
\label{eq: DiffEq v12 reduced}
\end{align}
Analogously, the single-massive differential equation becomes
%\begin{widetext}
\begin{align}
\frac{\partial I^{(1)}_{1,1,1}}{\partial v_{11}}&=
\left[\frac{(6-d)v_{23}}{2 (2v_{12}v_{13}-v_{11}v_{23})}+\frac{(d-5)v_{23}(2-v_{23})}{2 Y_{123}^{(1)}}\right]I^{(1)}_{1,1,1}\nonumber\\
&
+\frac{d-5}{2 Y_{123}^{(1)}(2v_{12}v_{13}-v_{11}v_{23})}\Big[
\Big((v_{12}+v_{13}-v_{23})(v_{12}+v_{23}-v_{13}-v_{12}v_{23})+Y_{123}^{(1)}\Big)I_{1,1,0}^{(1)}
\nonumber\\
&
+\Big((v_{12}+v_{13}-v_{23})(v_{13}+v_{23}-v_{12}-v_{13}v_{23})+Y_{123}^{(1)}\Big)I_{1,0,1}^{(1)}
\nonumber\\
&-(v_{12}+v_{13}-v_{23})v_{23}(2-v_{23})I_{0,1,1}^{(0)}
\Big]
%\nonumber\\
%&+\frac{(d-4)(1-v_{11})}{v_{11}(2v_{12}v_{13}-v_{11}v_{23})}I_{1,0,0}^{(1)}
+\frac{3-d}{v_{11}(2v_{12}v_{13}-v_{11}v_{23})}I_{0,0,0}^{(0)}\,.
\label{eq: DiffEq v11 reduced}
\end{align}
%\end{widetext}
We note that in both eqs.\,\eqref{eq: DiffEq v12 reduced} and \eqref{eq: DiffEq v11 reduced} all master integrals except for $I_{1,1,1}^{(0,1)}$ are known analytically in $d$ dimensions.

As discussed in section \ref{sec: Dimensional shift identity}, it is most useful to consider the differential equations in $d=6-2\eps$ dimensions.
The standard procedure to approach differential equations of this form is bringing them to canonical $\eps$-form, i.e. making the coefficient in front of $I_{1,1,1}$ proportional to $\eps$.
In the particular case at hand we can even make the entire coefficient disappear, leaving us with a right-hand side which is expressed solely in terms of known integrals
This is achieved by rescaling the integral to $J_{1,1,1} = f(v) I_{1,1,1}$.
Schematically writing the original eqs.\,\eqref{eq: DiffEq v12 reduced} and \eqref{eq: DiffEq v11 reduced} as
	\begin{align}
		\frac{\partial I_{1,1,1}}{\partial v} \,=\, c(v)\, I_{1,1,1} \,+\, r(v) \,,
	\end{align}
the corresponding differential equation for the rescaled integral is
	\begin{align}
		\frac{\partial J_{1,1,1}}{\partial v} \,= \left( \frac{1}{f(v)} \frac{\partial f(v)}{\partial v} \,+\, c(v) \right) J_{1,1,1} \,+\, f(v)\, r(v) \,.
	\end{align}
Choosing ${f(v) = f_0 \exp\left[-\int\dx v\, c(v)\right] }$, the term in parentheses vanishes.

Explicitly, we rescale the massless integral as
\begin{align}
J_{1,1,1}^{(0)}&\equiv(v_{12}v_{13}v_{23})^\eps \left(Y^{(0)}_{123}\right)^{1/2-\eps} I_{1,1,1}^{(0)}.
\end{align} 
The prefactor is chosen such that in the differential equation for $\partial/\partial v_{12} J_{1,1,1}^{(0)}$ the coefficient in front of $J_{1,1,1}^{(0)}$ becomes zero while simultaneously preserving the symmetry between variables.
For the single-massive integral we rescale as
\begin{align}
J_{1,1,1}^{(1)}&\equiv\left(\frac{v_{23}}{2}\right)^\eps (2v_{12}v_{13}-v_{11}v_{23})^\eps \left(Y^{(1)}_{123}\right)^{1/2-\eps}I_{1,1,1}^{(1)}\,,
\end{align}
making the coefficient in front of $J_{1,1,1}^{(1)}$ vanish in the differential equation for $\partial/\partial v_{11} J_{1,1,1}^{(1)}$ and also preserving the property that $J^{(1)}_{1,1,1}$ reduces to $J^{(0)}_{1,1,1}$ in the massless limit.
We note that while it is clear that with the prefactor chosen this way $J^{(1)}_{1,1,1}$ reduces to $J^{(0)}_{1,1,1}$ before the $\eps$-expansion, it also holds true in every order in the $\eps$-expansion since $I_{1,1,1}^{(0)}$ is finite around six dimensions.
All logarithmic divergences in the mass of the integral in $d=4-2\eps$ are already explicit in the two-denominator contributions of eq.\,\eqref{eq: I111 dimensional shift representation}.

Explicitly we find in the massless case
\begin{align}
&\frac{\partial J^{(0)}_{1,1,1}(\eps-1)}{\partial v_{12}}=
(v_{12}v_{13}v_{23})^\eps \left(Y^{(0)}_{123}\right)^{-1/2-\eps}
\frac{1-2\eps}{2\,v_{12}}
%\nonumber\\
%&\times
\Big[
v_{12}(4-v_{12}-v_{13}-v_{23})I^{(0)}_{1,1,0}(\eps-1)
\nonumber\\
&+(2-v_{13})(v_{23}-v_{12}-v_{13})I^{(0)}_{1,0,1}(\eps-1)
%\nonumber\\
%&\qquad
+(2-v_{23})(v_{13}-v_{12}-v_{23})I^{(0)}_{0,1,1}(\eps-1)
\Big].
\label{eq: d/dv12 closed form}
\end{align}
This constitutes a closed-form expression for ${\partial J^{(0)}_{1,1,1}(\eps-1)}/{\partial v_{12}}$ since the two-denominator integrals are known analytically in terms of hypergeometric functions as well as in terms of an all-order $\eps$-expansion.

For the derivative ${\partial J^{(1)}_{1,1,1}(\eps-1)}/{\partial v_{11}}$ we find a similar closed-form expression.
Here, it is convenient to express the right-hand side in terms of $I^{(1)}_{2,0,0}$ instead of $I^{(1)}_{1,0,0}$ to make spurious poles in $1/v_{11}$ cancel between coefficients, thus allowing for an integration starting at the massless point $v_{11}=0$.
Using the identity
\begin{align}
I_{1,0,0}(v_{11};\eps-1)=\frac{1}{2-2\eps}\left[(3-2\eps)I_{0,0,0}(\eps-1)-v_{11}\,I_{2,0,0}(v_{11};\eps-1)\right]
\end{align}
to substitute $I_{1,0,0}$ for $I_{2,0,0}$, the derivative of $J_{1,1,1}^{(1)}$ becomes
\begin{align}
\frac{\partial J_{1,1,1}^{(1)}(\eps-1)}{\partial v_{11}}=&2^{-1-\eps}v_{23}^\eps(2 v_{12}v_{13}-v_{11}v_{23})^{-1+\eps}\left(Y_{123}^{(1)}\right)^{-1/2-\eps}
\nonumber\\
&\times
\Big[(1-2\eps)\Big((v_{12}+v_{13}-v_{23})(v_{12}+v_{23}-v_{13}-v_{12}v_{23})+Y_{123}^{(1)}\Big)\,
I_{1,1,0}^{(1)}(\eps-1)
\nonumber\\
&
\quad+(1-2\eps)\Big((v_{12}+v_{13}-v_{23})(v_{13}+v_{23}-v_{12}-v_{13}v_{23})+Y_{123}^{(1)}\Big)\,
I_{1,0,1}^{(1)}(\eps-1)
\nonumber\\
&
\quad-(1-2\eps)
(v_{12}+v_{13}-v_{23}) v_{23}(2-v_{23})\,
I_{0,1,1}^{(0)}(\eps-1)
\nonumber\\
&
\quad-2(1-v_{11})Y_{123}^{(1)}\,
I_{2,0,0}^{(1)}(\eps-1)
-2(3-2\eps)Y_{123}^{(1)}\,
I_{0,0,0}^{(0)}(\eps-1)
\Big].
\label{eq: d/dv11 closed form}
\end{align}
All functions on the right-hand side of this equation are known to all orders in $\eps$.

\section{Integration for master integrals}
\label{sec: Integration for master integrals}
\subsection{Massless integral}
\begin{figure}
	\centering
	\includegraphics[width=1\textwidth]{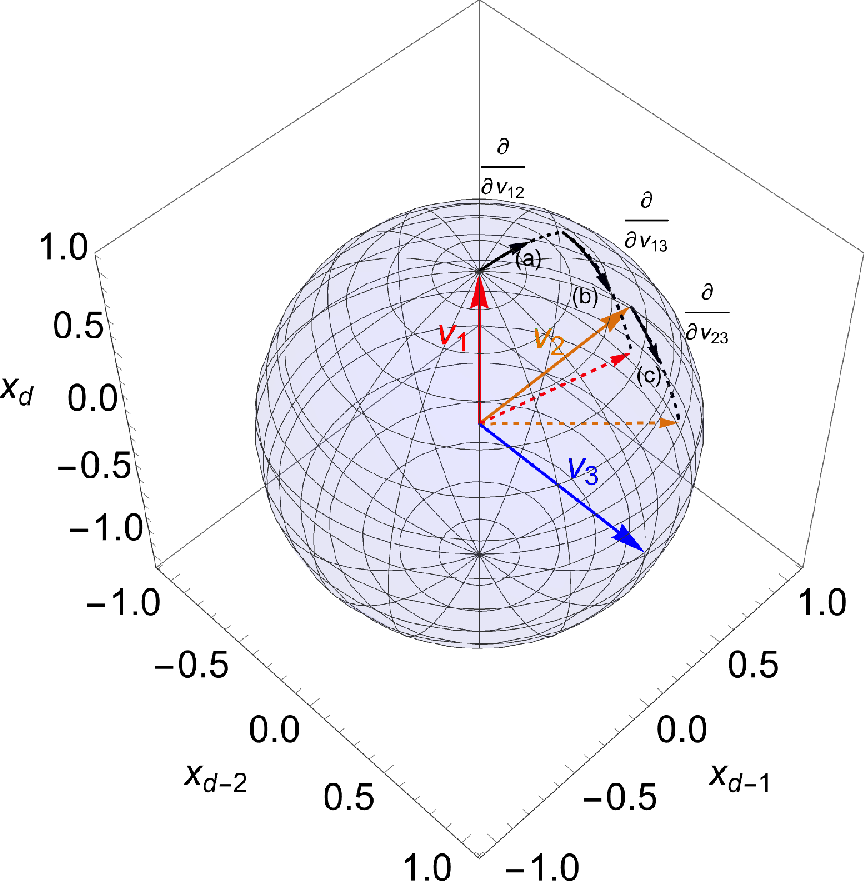}
	\caption{Illustration of the initial condition and the integration path for the massless three-denominator angular integral depending on vectors $v_{1,2,3}$. 
	The plot shows the slice $x_0=1$ of $d$-dimensional Minkowski space. The blue unit sphere depicts the intersection with the light-cone, where $v^2=0$.
	The symmetric configuration showcased by the solid arrows is used as the boundary value.
	In this configuration $I_{1,1,1}^{(0)}(1,1,1)=2\pi\catalan$, where $\catalan$ denotes Catalan's constant.
	The dashed black lines showcase the integration path.
	First, we rotate $v_1$ along path $(a)$, which is in the direction of $\partial/\partial v_{12}$.
	The axis of rotation is along $v_3$. 
	Second, we rotate $v_1$ along path $(b)$, which is in the direction of $\partial/\partial v_{13}$. 
	The axis of rotation is along $v_2$.
	Third, we rotate $v_2$ along path $(c)$, which is in the direction of $\partial/\partial v_{23}$, reaching the final configuration with general $v_{12}$, $v_{13}$, $v_{23}$.
	The axis of rotation is the rotated $v_1$ shown in dashed red.
	Note that the volume of the tetrahedron spanned by the vectors $v_{1,2,3}$ in this subspace is proportional to $\sqrt{Y_{123}^{(0)}}$\,.}
	\label{fig:Boundary value}
	\end{figure}

To obtain the order $\eps$ for $I_{1,1,1}^{(0)}$ in $d=4-2\eps$ dimensions we only need the leading term of eq.~\eqref{eq: d/dv12 closed form} which is of order $\eps^0$.
Hence, we can set $\eps=0$ in this equation and plug in the massless two-denominator integral in $d=6$ dimensions.
Here we use the dimensionally-shifted massless two-denominator master integral given in eq.\,\eqref{eq: I11 massless d=6} of appendix \ref{app: Expansion of dimensionally shifted two-denominator integrals}
leading to
%\begin{widetext}
\begin{align}
\frac{\partial J^{(0)}_{1,1,1}(-1)}{\partial v_{12}}&=\frac{\pi}{\sqrt{Y_{123}^{(0)}}}\left[
\log\frac{v_{13}v_{23}}{4}
-\log\frac{v_{12}}{2}
+\frac{(v_{13}-v_{23})\log\frac{v_{13}}{v_{23}}}{v_{12}}
-\frac{(2-v_{13}-v_{23})\log\frac{v_{12}}{2}}{2-v_{12}}
\right]
\nonumber\\
&\equiv j^{(0)}(v_{12},v_{13},v_{23})\,.
\label{eq: Integration kernel massless order eps}
\end{align}
%\end{widetext}
To obtain $J^{(0)}_{1,1,1}$ at a general kinematic point, we may choose the integration path $(1,1,1)\rightarrow (1,1,v_{23})\rightarrow (1,v_{13},v_{23})\rightarrow (v_{12},v_{13},v_{23})$ with boundary condition at the symmetric point $v_{12}=v_{13}=v_{23}=1$. 
The configuration of vectors used for the boundary value and the integration path are depicted in figure \ref{fig:Boundary value}.
By explicit calculation, the necessary boundary value is found to be
\begin{align}
J^{(0),d=6}_{1,1,1}(1,1,1)=I^{(0),d=6}_{1,1,1}(1,1,1)=2\pi\,\catalan \,,
\label{eq: Boundary value}
\end{align}
where $\catalan=\sum_{n=0}^\infty\frac{(-1)^n}{(2n+1)^2}$ denotes Catalan's constant\footnote{Obviously, $\catalan$ is a suitable name for an integration constant.}.
Details of the calculation can be found in appendix \ref{app: Calculation of the boundary value for the massless master integral}.

Utilizing the symmetry of the massless integral it is
\begin{align}
J^{(0),d=6}_{1,1,1}(v_{12},v_{13},v_{23})=&\int_1^{v_{12}}\!\!\!\!\dx v\,j^{(0)}(v,v_{13},v_{23})
%\nonumber\\
%&\qquad
+\int_1^{v_{13}}\!\!\!\!\dx v\, j^{(0)}(v,1,v_{23})
\nonumber\\+&\int_1^{v_{23}}\!\!\!\!\dx v\, j^{(0)}(v,1,1)
%\nonumber\\
%&\qquad
+J^{(0),d=6}_{1,1,1}(1,1,1)\,.
\label{eq: Integration path}
\end{align}
We note that alternatively $J_{1,1,1}^{(0)}$ is also fixed by being the unique symmetric solution to eq.\,\eqref{eq: Integration kernel massless order eps} with the boundary condition \eqref{eq: Boundary value}.

To evaluate the integrals from \eqref{eq: Integration path} in terms of (generalized) polylogarithms, the only remaining obstacle is the presence of the square root $\sqrt{Y_{123}^{(0)}}$.
A general algorithmic approach to rationalize roots of this kind has been presented in \cite{Besier:2018}.
In the present case the required change of variables is $v_{12}\rightarrow u$ with
	\begin{align}
		v_{12}=\frac{v_+ +v_- u^2}{1+u^2}\quad\text{and Jacobian }\frac{\dx v_{12}}{\dx u}=-\frac{2u(v_+-v_-)}{(1+u^2)^2}\,,
	\end{align}
where $v_\pm$ are defined such that they factorize the square root, i.e. $Y_{123}^{(0)}=(v_+-v_{12})(v_{12}-v_-)$, explicitly
	\begin{align}
		v_\pm\equiv v_{13}+v_{23}-v_{13}v_{23}\pm\sqrt{v_{13}(2-v_{13})v_{23}(2-v_{23})}\,.
	\end{align}
The square root of the Gram determinant becomes in the new variable $u$
\begin{align}
\sqrt{Y_{123}^{(0)}}= \frac{2u}{1+u^2}\,\sqrt{v_{13}(2-v_{13})v_{23}(2-v_{23})}\,.
\end{align}
The integrals resulting from this change of variables take the form of $J_0(\delta)$ and $J_2(\beta,\delta)$ defined in appendix \ref{app: Useful integrals for the calculation of the master integrals}.
After considerable simplifications between the terms we obtain the result given in section \ref{sec: Results Massless integral}.

\subsection{Single-massive integral}
To obtain the order $\eps$ for $I_{1,1,1}^{(1)}$ in $d=4-2\eps$ dimensions we only need the leading $\eps^0$ term of eq.~\eqref{eq: d/dv11 closed form}.
Hence, we can again simply put $\eps=0$ in this equation and plug in the known integrals with two and fewer denominators in $d=6$. Besides eq.\eqref{eq: I11 massless d=6} we use the dimensionally-shifted two-denominator integrals from appendix \ref{app: Expansion of dimensionally shifted two-denominator integrals} leading to
\begin{align}
&\frac{\partial J_{1,1,1}^{(1)}(-1)}{\partial v_{11}}=
\frac{\pi}{\sqrt{Y_{123}^{(1)}}}
\left[
\frac{ (v_{12}-1) (v_{12}+v_{23}-v_{13}-v_{12} v_{23})\log
   \!\left(\frac{1+\sqrt{1-v_{11}}}{1-\sqrt{1-v_{11}}}\right)}{2 (v_{12}(2-v_{12})-v_{11})\sqrt{1-v_{11}}}
   \right.
   \nonumber\\
   &+\frac{(v_{13}-1) (v_{13}+v_{23}-v_{12}-v_{13}v_{23})\log
   \!\left(\frac{1+\sqrt{1-v_{11}}}{1-\sqrt{1-v_{11}}}\right)}{2
   (v_{13}(2-v_{13})-v_{11})\sqrt{1-v_{11}}}
%   \nonumber\\
%   &
   +\frac{(v_{12}+v_{23}-v_{13}-v_{12}v_{23}) \log \frac{v_{11}}{v_{12}^2}}{2 (v_{12}(2-v_{12})-v_{11})}
   \nonumber\\
   &
   +\frac{(v_{13}+v_{23} -v_{12}-v_{13} v_{23}) \log
   \frac{v_{11}}{v_{13}^2}}{2 (v_{13}(2-v_{13})-v_{11})}
%   \nonumber\\
%   &+\frac{(-v_{13}-v_{12} (v_{23}-1)+v_{23}) \log v_{12}}{v_{11}+(v_{12}-2) v_{12}}+\frac{(-v_{12}-v_{13} (v_{23}-1)+v_{23})
%   \log v_{13}}{v_{11}+(v_{13}-2) v_{13}}
%\nonumber\\   
%   &
\left.
+\frac{(v_{12}+v_{13}-v_{23}) v_{23} \log\!\left( \frac{v_{11}v_{23}}{2v_{12}v_{13}}\right)}{2 v_{12} v_{13}-v_{11} v_{23}}
\right]\equiv j^{(1)}(v_{11})\,.
\label{eq: Integration kernel massive order eps}
\end{align}
Integrating the right-hand side starting at $v_{11}=0$ we can use the massless integral $J^{(0),d=6}_{1,1,1}$ as a boundary value,
\begin{align}
J_{1,1,1}^{(1),d=6}(v_{11})=J_{1,1,1}^{(0),d=6}+\int_0^{v_{11}}\dx v\,j^{(1)}(v)\,.
\end{align}
Note that when rescaling back from $J_{1,1,1}$ to $I_{1,1,1}$ the massless boundary term receives a $v_{11}$-dependent prefactor, namely
\begin{align}
I_{1,1,1}^{(1),d=6}=\frac{J_{1,1,1}^{(1),d=6}}{\sqrt{Y_{123}^{(1)}}}=\sqrt{\frac{Y_{123}^{(0)}}{Y_{123}^{(1)}}}\,I_{1,1,1}^{(0),d=6}+\frac{1}{\sqrt{Y_{123}^{(1)}}}\int_0^{v_{11}}\dx v\,j^{(1)}(v)\,.
\end{align}

To express the remaining integrals in terms of polylogarithms we again make use of suitable rationalizations.
For the terms where only the square root $\sqrt{Y_{123}^{(1)}}$ is present, such a change of variables $v_{11}\rightarrow t$ is given by
\begin{align}
v_{11}=\frac{t^2 w_{23}-1}{t^2}\text{ with Jacobian }\frac{\dx v_{11}}{\dx t}=\frac{2}{t^3}\,,
\end{align}
where $w_{23}=\frac{Y_{123}^{(0)}}{v_{23}(2-v_{23})}$.
In the variable $t$, the square root becomes
\begin{align}
\sqrt{Y^{(1)}_{123}}=\frac{\sqrt{v_{23}(2-v_{23})}}{t}\,.
\end{align}
For the integrals additionally containing the square root $\sqrt{1-v_{11}}$, a change of variables $v_{11}\rightarrow s$ that simultaneously rationalizes both roots can be found using the approach of \cite{Besier:2018} and is given by
\begin{align}
v_{11}=\frac{Y_{123}^{(1)} s^4 +2 s^2(Y_{123}^{(1)}-2)+Y_{123}^{(1)}}{(1-s)^2(1+s)^2}\text{ with Jacobian }\frac{\dx v_{11}}{\dx s}=\frac{8 s (1+s^2)(w_{23}-1)}{(1-s^2)^3}\,.
\end{align}
In terms of this new variable $s$ the square roots become
\begin{align}
\sqrt{Y^{(1)}_{123}}=\frac{2s}{(1-s)(1+s)}\,\sqrt{v_{23}(2-v_{23})}\,\sqrt{1-w_{23}}\;\;\text{and}\;\;
\sqrt{1-v_{11}}=\frac{1+s^2}{(1-s)(1+s)}\,\sqrt{1-w_{23}}\,.
\end{align}

The integrals resulting from the change of variables take the form of $J_0(\delta)$ and $J_1(\alpha,\delta)$ defined in appendix \ref{app: Useful integrals for the calculation of the master integrals}.
After considerable simplifications between the terms we obtain the result given in section \ref{sec: Results Massive integral}.

\section{Results}
\label{sec: Results}
In this section we present the $\eps$-expansion for the massless, single-, double-, and triple-massive master integrals $I^{(m)}_{1,1,1}$ including order $\eps$ terms.
In combination with the recursion relations listed in appendix \ref{app: Recursion relations} these give the $\eps$-expansion of all three-denominator angular integrals $I^{(m)}_{j_1,j_2,j_3}$ with integer-valued propagator powers $j_i$.
The results are presented in a form that is manifestly real-valued for time- and light-like real vectors $v_i$ as will be the case in most applications.
A detailed study of the non-trivial branch-cut structure, which arises outside this region, is beyond the scope of this paper.
\subsection{Massless integral}
\label{sec: Results Massless integral}
For the massless three-denominator integral we find the $\eps$-expansion
%\begin{widetext}
\begin{align}
&I_{1,1,1}^{(0)}(v_{12},v_{13},v_{23};\eps)=-\frac{\pi}{\eps}\frac{(v_{12}+v_{13}+v_{23})}{v_{12}v_{13}v_{23}}
\nonumber\\
&+\frac{\pi}{v_{12}v_{13}v_{23}}
\left[
(v_{13}+v_{23}-v_{12})\log\frac{v_{12}}{2}
+(v_{12}+v_{23}-v_{13})\log\frac{v_{13}}{2}
+(v_{12}+v_{13}-v_{23})\log\frac{v_{23}}{2}
\right]
\nonumber\\
&+\frac{\pi\,\eps}{v_{12}v_{13}v_{23}}
\left[
(v_{13}+v_{23}-v_{12})\,\dilog\!\left(1-\frac{2}{v_{12}}\right)
+(v_{12}+v_{23}-v_{13})\,\dilog\!\left(1-\frac{2}{v_{13}}\right)
\right.
\nonumber\\
&\left.\phantom{+\frac{\pi\,\eps}{v_{12}v_{13}v_{23}}\Big[}+(v_{12}+v_{13}-v_{23})\,\dilog\!\left(1-\frac{2}{v_{23}}\right)
+ \sqrt{Y_{123}^{(0)}} f^{(0)}_Y(v_{12},v_{13},v_{23})
\right]+\mathcal{O}(\eps^2) ,
\label{eq: massless result}
\end{align}
where
\begin{align}
f^{(0)}_Y(v_{12},v_{13},v_{23})&=\frac{\sqrt{Y^{(0)}_{123}}}{\pi} I_{1,1,1}^{(0),d=6}=\frac{J_{1,1,1}^{(0)}(-1)}{\pi}
 \nonumber\\
&=\clausen\!\left(2\theta_1^{(0)}\right)+\clausen\!\left(2\theta_2^{(0)}\right)+\clausen\!\left(2\theta_3^{(0)}\right)
+\clausen\!\left(2\theta_{12}^{(0)}\right)
\nonumber\\
&+\clausen\!\left(2\theta_{13}^{(0)}\right)
+\clausen\!\left(2\theta_{23}^{(0)}\right)-\clausen\!\left(2\theta_{123}^{(0)}\right)
\label{eq: result fX0}
\end{align}
%\end{widetext}
with the abbreviations 
\begin{align}
&\theta_i^{(m)}=\arctan\frac{\sqrt{Y_{123}^{(m)}}}{v_{ij}+v_{ik}-v_{jk}-v_{ij}v_{ik}}\,,\quad
\theta_{ij}^{(m)}=\arctan\frac{\sqrt{Y_{123}^{(m)}}}{v_{ik}+v_{jk}-v_{ij}}\,,\nonumber\\
&
\theta_{ijk}^{(m)}=\arctan\frac{\sqrt{Y_{123}^{(m)}}}{4-v_{ij}-v_{ik}-v_{jk}}\,,
\label{eq: Theta angles}
\end{align}
where $i,j,k\in\lbrace1,2,3\rbrace$ always denote pairwise distinct indices.
$\clausen$ is the Clausen function recalled in appendix \ref{app: Clausen function}.
Note that $f_{Y}^{(0)}(v_{12},v_{13},v_{23})$ is manifestly symmetric in its arguments.
We recall our observation from section \ref{sec: Differential operators for angular integrals} that $\sqrt{Y_{123}^{(0)}}$\,, which is given in eq.\,\eqref{eq: massless X}, is proportional to the volume of the tetrahedron spanned by $\vec{v}_{1,2,3}$ and hence vanishes if and only if the vectors $v_{1,2,3}$ become linearly dependent.
In this case, the result reduces to a sum of two-denominator integrals matching the result one obtains by partial fractioning.
The pole and finite parts are in agreement with the results from \cite{Somogyi:2011}, which have been calculated by Mellin-Barnes techniques\,\footnote{Beware of the difference in normalization of the scalar products, to compare with \cite{Somogyi:2011} one needs to set $v_{ij}\rightarrow 2 v_{ij}$ in our results.}.

\begin{figure}
\centering
\includegraphics[width=0.55\textwidth]{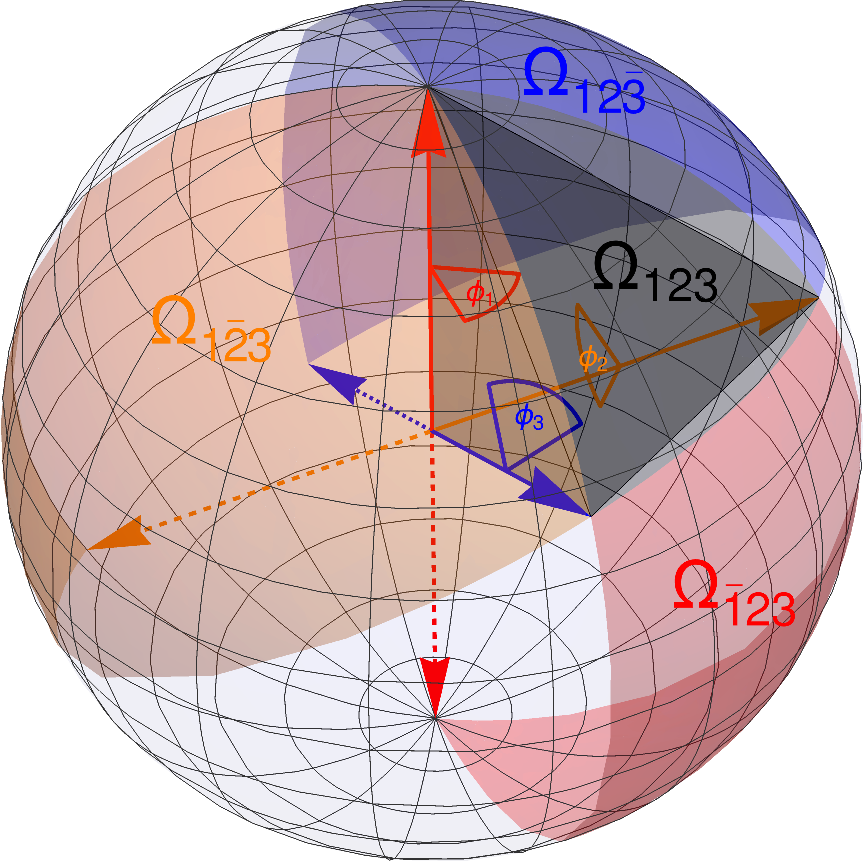}
\caption{Visualization of the geometrical quantities appearing in the result \eqref{eq: Geometrical interpretation of massless integral} for the massless integral in six dimensions.
Analogous to figure \ref{fig:Boundary value}, the picture shows the slice $x_0=1$ of six-dimensional Minkowski space, specifically the three-dimensional subspace defined by $\vec{v}_{1}$, $\vec{v}_{2}$, $\vec{v}_{3}$.
These vectors span a tetrahedron with volume $\mathbb{V}(\vec{v}_1,\vec{v}_2,\vec{v}_3)$ shaded in gray. 
Note that the factor of $6$ in eq.\,\eqref{eq: Geometrical interpretation of massless integral} is not from $d=6$ but originates from eq.\,\eqref{eq: Y123 geometrically}.
The dihedral angle between the faces that meet at $\vec{v}_i$ is denoted by $\phi_i$, i.e. e.g. $\phi_1$ is the angle between the face spanned by $(\vec{v}_1,\vec{v}_2)$ and $(\vec{v}_1,\vec{v}_3)$.
They contribute with $\clausen(2\phi_i)$ in eq.\,\eqref{eq: Geometrical interpretation of massless integral}.
The $\Omega$ are solid angles, i.e. areas of triangles on the the unit sphere.
The red area $\Omega_{\bar{1}23}$ has corners at $(-\vec{v}_1,\vec{v}_2,\vec{v}_3)$, the yellow area $\Omega_{1\bar{2}3}$ has corners at $(\vec{v}_1,-\vec{v}_2,\vec{v}_3)$, and the blue area $\Omega_{\bar{1}23}$ has corners at $(\vec{v}_1,\vec{v}_2,-\vec{v}_3)$.
They contribute with $\clausen(\Omega)$ in eq.\,\eqref{eq: Geometrical interpretation of massless integral}.
The black area $\Omega_{123}$ is the solid angle covered by $(\vec{v}_1,\vec{v}_2,\vec{v}_3)$ and contributes negatively with $-\clausen(\Omega)$ in eq.\,\eqref{eq: Geometrical interpretation of massless integral}.}
\label{fig: Geometrical interpretation of massless integral}
\end{figure}

Upon closer inspection of eq.\,\eqref{eq: result fX0} we notice that the massless angular integral in six dimensions can be satisfyingly interpreted in geometrical terms.
Considering the three-dimensional tetrahedron spanned by the unit vectors $(\vec{v}_1,\vec{v}_2,\vec{v}_3)$, we denote with $\phi_i$ the dihedral angle between the faces that meet at the edge $\vec{v}_{i}$.
Calculating these angle as the angle between the vectors normal to the respective faces we find
\begin{align}
\tan \phi_i=\frac{|(\vec{v}_i\times\vec{v}_j)\times(\vec{v}_i\times\vec{v}_k)|}{(\vec{v}_i\times\vec{v}_j)\cdot (\vec{v}_i\times\vec{v}_k)}=\frac{|\det(\vec{v}_i,\vec{v}_j,\vec{v}_k)|}{\vec{v}_j\!\cdot\!\vec{v}_k-\vec{v}_i\!\cdot\!\vec{v}_j\,\vec{v}_i\!\cdot\!\vec{v}_k}=\tan\theta_i^{(0)}\,.
\end{align}
Further we write $\Omega_{123}$ for the solid angle covered by the vectors $(\vec{v}_1,\vec{v}_2,\vec{v}_3)$ -- in other words, the area of the triangle on the unit sphere with edges at $(\vec{v}_1,\vec{v}_2,\vec{v}_3)$.
Flipping one vector $\vec{v}_i\rightarrow -\vec{v}_i$, we define the additional solid angles $\Omega_{\bar{1}23}$, $\Omega_{1\bar{2}3}$ and, $\Omega_{12\bar{3}}$ where the flipped vector in each case is marked with a bar.
Using Euler's identity for spherical triangles from §23 of \cite{Euler:1781}, given in modern notation in  \cite{Eriksson:1990},
\begin{align}
\tan\frac{\Omega_{123}}{2}=\frac{|\det(\vec{v}_1,\vec{v}_2,\vec{v}_3)|}{1+\vec{v}_1\!\cdot\!\vec{v}_2+\vec{v}_1\!\cdot\!\vec{v}_3+\vec{v}_2\!\cdot\!\vec{v}_3},
\end{align}
we find 
\begin{align}
2\theta_{123}^{(0)}=\Omega_{123}\,,\quad
2\theta_{12}^{(0)}=\Omega_{12\bar{3}}\,,\quad
2\theta_{13}^{(0)}=\Omega_{1\bar{2}3}\,,\quad\text{and}\quad
2\theta_{23}^{(0)}=\Omega_{\bar{1}23}\,.
\end{align} 

Hence the massless three-denominator angular integral in six dimensions can be presented in the form
\begin{align}
I_{1,1,1}^{(0),d=6}=\frac{\pi}{6\mathbb{V}(\vec{v}_1,\vec{v}_2,\vec{v}_3)}&\Big[\clausen(2\phi_1)+\clausen(2\phi_2)+\clausen(2\phi_3)+\clausen(\Omega_{\bar{1}23})
\nonumber\\&+\clausen(\Omega_{1\bar{2}3})+\clausen(\Omega_{12\bar{3}})-\clausen(\Omega_{123})\Big],
\label{eq: Geometrical interpretation of massless integral}
\end{align}
where the geometric meaning of each object is visualized in figure \ref{fig: Geometrical interpretation of massless integral}.
Curiously, we observe that the Clausen function is related to the volume of hyperbolic tetrahedra (see appendix \ref{app: Clausen function}), while the arguments in eq.\,\eqref{eq: Geometrical interpretation of massless integral} are all given in relation to Euclidean tetrahedra.
With dihedral angles, solid angles, and the Clausen function eq.\,\eqref{eq: Geometrical interpretation of massless integral} combines quantities from Euclidean, spherical, and hyperbolic geometry.

\subsection{Single-massive integral}
\label{sec: Results Massive integral}
For the single-massive three-denominator integral we find the $\eps$-expansion
%\begin{widetext}
\begin{align}
&I_{1,1,1}^{(1)}(v_{12},v_{13},v_{23},v_{11};\eps)=-\frac{\pi}{\eps}\frac{(v_{12}+v_{13})}{v_{12}v_{13}v_{23}}
\nonumber\\
&+\frac{\pi}{v_{12}v_{13}v_{23}(2 v_{12}v_{13}-v_{11}v_{23})}\Big[
v_{13}(v_{11}v_{23}+v_{12}^2-v_{12}v_{13}-v_{12}v_{23})\log\frac{v_{11}}{v_{12}^2}\nonumber\\
&\qquad
+v_{12}(v_{11}v_{23}+v_{13}^2-v_{12}v_{13}-v_{13}v_{23})\log\frac{v_{11}}{v_{13}^2}
+2v_{12}v_{13}(v_{12}+v_{13}-v_{23})\log\frac{v_{23}}{2}\Big]
\nonumber\\
&+\frac{2\pi\,\eps}{v_{12}v_{13}v_{23}(2 v_{12}v_{13}-v_{11}v_{23})}\left[\vphantom{\frac{•}{•}}\right.v_{12}v_{13}(v_{12}+v_{13}-v_{23})\,
\dilog\!\left(1-\frac{2}{v_{23}}\right)
\nonumber\\
&\qquad
+v_{13}(v_{11}v_{23}+v_{12}^2-v_{12}v_{13}-v_{12}v_{23})
\nonumber\\&\qquad\quad\times
\left(\dilog\!\left(1-\frac{v_{12}}{1+\sqrt{1-v_{11}}}\right)+
\dilog\!\left(1-\frac{v_{12}}{1-\sqrt{1-v_{11}}}\right)
+\frac{1}{4}\log\frac{v_{11}}{v_{12}^2}\right)
\nonumber\\
&\qquad
+v_{12}(v_{11}v_{23}+v_{13}^2-v_{12}v_{13}-v_{13}v_{23})
\nonumber\\&\qquad\quad\times\left(\dilog\!\left(1-\frac{v_{13}}{1+\sqrt{1-v_{11}}}\right)+
\dilog\!\left(1-\frac{v_{13}}{1-\sqrt{1-v_{11}}}\right)
+\frac{1}{4}\log\frac{v_{11}}{v_{13}^2}\right)
\nonumber\\
&\qquad+v_{12}v_{13}\sqrt{Y_{123}^{(1)}}f_Y^{(1)}(v_{12},v_{13},v_{23},v_{11})
\left.\vphantom{\frac{ }{ }}\right] ,
\label{eq: single-massive result}
\end{align}
%\end{widetext}
where
\begin{align}
f_Y^{(1)}&(v_{12},v_{13},v_{23},v_{11})=\frac{\sqrt{Y_{123}^{(1)}}}{\pi}I_{1,1,1}^{(1),d=6}=\frac{J_{1,1,1}^{(1)}(-1)}{\pi}=f_Y^{(0)}+\int_0^{v_{11}}\dx v\,j^{(1)}(v)\nonumber\\
=&\,
   \clausen\!\left(2 \theta_1^{(0)}\right)
   +\clausen\!\left(2 \theta_2^{(1)}\right)
   +\clausen\!\left(2 \theta_3^{(1)}\right)
   +\clausen\!\left(2 \theta_{12}^{(0)}\right)
   +\clausen\!\left(2 \theta_{13}^{(0)}\right)
        \nonumber\\
& 
   +2 \clausen\!\left(2 \theta_{23}^{(1)}\right)
   -\clausen\!\left(2 \theta_{23}^{(0)}\right)
   -\clausen\!\left(2 \theta_{123}^{(0)}\right)
      \nonumber\\
&
+\frac{1}{2} \clausen\!\left(2 \theta_2^{(1)}+ 2\theta_2^{(0)}\right)
+\frac{1}{2}\clausen\!\left(2 \theta _2^{(1)}-2\theta_2^{(0)}\right)
   -\frac{1}{2} \clausen\!\left(4 \theta_2^{(1)}\right)
   \nonumber\\
&
+\frac{1}{2}\clausen\!\left(2 \theta _3^{(1)}+2\theta_3^{(0)}\right)
+\frac{1}{2} \clausen\!\left(2 \theta_3^{(1)}-2\theta _3^{(0)}\right)
-\frac{1}{2} \clausen\!\left(4 \theta_3^{(1)}\right)
\nonumber\\
&  
   +\clausen\!\left(2\theta_{23}^{(1)}+2\theta_{23}^{(0)}\right)
   +\clausen\!\left(2 \theta_{23}^{(1)}-2\theta _{23}^{(0)}\right)
   -\clausen\!\left(4 \theta_{23}^{(1)}\right)
         \nonumber\\
&   
+\frac{1}{2} \clausen\!\left(2\phi_{12,+}^{(1)}+2\phi_{12,+}^{(0)}\right)
   +\frac{1}{2} \clausen\!\left(2\phi_{12,+}^{(1)}-2\phi_{12,+}^{(0)}\right) 
    +\frac{1}{2} \clausen\!\left(2\phi_{12,-}^{(1)}+2\phi_{12,-}^{(0)}\right)
          \nonumber\\
&
   +\frac{1}{2} \clausen\!\left(2\phi_{12,-}^{(1)}-2\phi_{12,-}^{(0)}\right)
-\frac{1}{2} \clausen\!\left(2\phi_{12,+}^{(1)}+2\phi_{12,-}^{(0)}\right)
-\frac{1}{2} \clausen\!\left(2\phi_{12,+}^{(1)}-2\phi_{12,-}^{(0)}\right)
      \nonumber\\
&
-\frac{1}{2} \clausen\!\left(2\phi_{12,-}^{(1)}+2\phi
   _{2,+}^{(0)}\right)
-\frac{1}{2}\clausen\!\left(2 \phi_{12,-}^{(1)}-2\phi_{12,+}^{(0)}\right)
\nonumber\\
&
   +\clausen\!\left(2 \phi_{12,+}^{(0)}+2\phi_{12,-}^{(0)}\right)
   -\frac{1}{2} \clausen\!\left(4\phi_{12,+}^{(0)}\right)-\frac{1}{2} \clausen\!\left(4\phi_{12,-}^{(0)}\right)
            \nonumber\\
         &
  +\frac{1}{2} \clausen\!\left(2\phi_{13,+}^{(0)}+2\phi_{13,+}^{(1)}\right)
     +\frac{1}{2} \clausen\!\left(2\phi_{13,+}^{(1)}-2\phi_{13,+}^{(0)}\right)
   +\frac{1}{2} \clausen\!\left(2\phi_{13,-}^{(1)}+2\phi_{13,-}^{(0)}\right)
         \nonumber\\
&
   +\frac{1}{2} \clausen\!\left(2\phi_{13,-}^{(1)}-2\phi_{13,-}^{(0)}\right)
   -\frac{1}{2} \clausen\!\left(2\phi_{13,+}^{(1)}+2\phi_{13,-}^{(0)}\right)
      -\frac{1}{2} \clausen\!\left(2\phi_{13,+}^{(1)}-2\phi_{13,-}^{(0)}\right)
            \nonumber\\
&
      -\frac{1}{2} \clausen\!\left(2\phi_{13,-}^{(1)}+2\phi_{13,+}^{(0)}\right)
   -\frac{1}{2} \clausen\!\left(2 \phi_{13,-}^{(1)}-2\phi_{13,+}^{(0)}\right)
      \nonumber\\
&
   +\clausen\!\left(2 \phi_{13,+}^{(0)}+2\phi_{13,-}^{(0)}\right)
   -\frac{1}{2} \clausen\!\left(4\phi_{13,+}^{(0)}\right)-\frac{1}{2} \clausen\!\left(4\phi_{13,-}^{(0)}\right),
\label{eq: result fX1}
\end{align}
with the abbreviations from \eqref{eq: Theta angles} and
\begin{align}
\phi_{1j,\pm}^{(0)}&=\arctan\!\left(\frac{\sqrt{Y_{123}^{(0)}}}{v_{1j}+v_{jk}-v_{1k}-v_{1j}v_{jk}}\,\frac{1-v_{1j}+\sqrt{1-w_{jk}}}{1\pm \sqrt{1-w_{jk}}}\right),
\\
\phi_{1j,\pm}^{(1)}&=\arctan\!\left(\frac{\sqrt{Y_{123}^{(1)}}}{v_{1j}+v_{jk}-v_{1k}-v_{1j}v_{jk}}\,\frac{1-v_{1j}+\sqrt{1-w_{jk}}}{\sqrt{1-v_{11}}\pm \sqrt{1-w_{jk}}}\right),
\label{eq: Phi angles}
\end{align}
where ${j,k}\in \lbrace 2,3\rbrace$ with $j\neq k$ and
\begin{align}
w_{jk}=\frac{Y_{123}^{(0)}}{v_{jk}(2-v_{jk})}\,.
\end{align}
The quantities $Y_{123}^{(0,1)}$ are given in eqs.\,\eqref{eq: massless X} respectively \eqref{eq: single-massive X}.
As in the massless case, the result simplifies to a sum of two-denominator integrals in the case where the vectors $\vec{v}_{1,2,3}$ become linearly dependent.
The pole part is identical to the result found in \cite{Smirnov:2024}, while the finite part agrees in the limit of small mass $v_{11}$ with the result calculated in \cite{Smirnov:2024} via expansion by regions.

\subsection{Double- and triple-massive integral}
The $\eps$-expansions of the double- and triple-massive integral could be obtained by plugging the massless and single-massive results from eqs.\eqref{eq: massless result} and \eqref{eq: single-massive result} into eqs.\eqref{eq: reduction of two-mass integral} and \eqref{eq: reduction of three-mass integral}.
However, it is much more economical to first use eq.\,\eqref{eq: I111 dimensional shift representation} to extract the part that is expressible in terms of known two-denominator integrals.
Subsequently eqs.\eqref{eq: reduction of two-mass integral} and \eqref{eq: reduction of three-mass integral} need to be applied only to $I_{1,1,1}^{d=6}$.
By absorbing a factor of $\sqrt{Y_{123}}$, we can apply the mass splitting in the form of \eqref{eq: reduction of two-mass integral simplified} respectively \eqref{eq: reduction of three-mass integral simplified}.
Therefore the order $\eps$ term in the double- respectively triple-massive case is conveniently expressed in terms of the functions
\begin{align}
f_Y^{(2)}(v_{12},v_{13},v_{23},v_{11},v_{22})=&
f_Y^{(0)}\left(v_{3(12)},v_{3(21)},v_{(12)(21)}\right)
-f_Y^{(1)}\left(v_{13},v_{1(12)},v_{3(12)},v_{11}\right)
\nonumber\\
&
-f_Y^{(1)}\left(v_{23},v_{2(21)},v_{3(21)},v_{22}\right),
\label{eq: fY2}
\end{align}
respectively
\begin{align}
f_Y^{(3)}&(v_{12},v_{13},v_{23},v_{11},v_{22},v_{33})=
f_Y^{(1)}\left(v_{3(12)},v_{3(21)},v_{(12)(21)},v_{33}\right)
\nonumber\\
&
-f_Y^{(0)}\left(v_{(12)(13)},v_{(12)(31)},v_{(13)(31)}\right)
+f_Y^{(1)}\left(v_{1(12)},v_{1(13)},v_{(12)(13)},v_{11}\right)
\nonumber\\
&+f_Y^{(1)}\left(v_{3(12)},v_{3(31)},v_{(12)(31)},v_{33}\right)
-f_Y^{(0)}\left(v_{(21)(23)},v_{(21)(32)},v_{(23)(32)}\right)
\nonumber\\
&+f_Y^{(1)}\left(v_{2(21)},v_{2(23)},v_{(21)(23)},v_{22}\right)
+f_Y^{(1)}\left(v_{3(21)},v_{3(32)},v_{(21)(32)},v_{33}\right).
\label{eq: fY3}
\end{align}
The necessary massless and single-massive functions $f_Y^{(0)}$ and $f_Y^{(1)}$ are given in eqs.\,\eqref{eq: result fX0} and \eqref{eq: result fX1}.
For compact notation we introduce the further functions
\begin{align}
f_1^{(0)}(v_{ij})&\equiv 2\,\dilog\!\left(1-\frac{2}{v_{ij}}\right),\\
f_1^{(1)}(v_{ij},v_{ii})&\equiv-2\,\dilog\!\left(1-\frac{v_{ij}}{1+\sqrt{1-v_{ii}}}\right)-2\dilog\!\left(1-\frac{v_{ij}}{1-\sqrt{1-v_{ii}}}\right)-\frac{1}{2}\log^2\!\left(\frac{v_{ii}}{v_{ij}^2}\right),\\
f_1^{(2)}(v_{ij},v_{ii},v_{jj})&\equiv f_1^{(0)}(v_{(ij)(ji)})-f_1^{(1)}(v_{i(ij)},v_{ii})-f_1^{(1)}(v_{j(ji)},v_{jj})\,.
\end{align}
With these abbreviations in place, the $\eps$-expansion of the double-massive three-denominator integral is given by
\begin{align}
I_{1,1,1}^{(2)}&(v_{12},v_{13},v_{23},v_{11},v_{22};\eps)=-\frac{\pi}{\eps}\frac{1}{v_{13}v_{23}}
+\frac{\pi}{2 v_{12} v_{23} v_{13}-v_{11} v_{23}^2-v_{22} v_{13}^2}
\nonumber\\
&\times
\left[
\Big(v_{13}-v_{12}-v_{23}+\frac{v_{11}
   v_{23}}{v_{13}}\Big) \log\frac{v_{11}}{v_{13}^2}
   \right.
+
   \Big(v_{23}-v_{12}-v_{13}+\frac{v_{22}v_{13}}{v_{23}}
   \Big) \log\frac{v_{22}}{v_{23}^2}
   \nonumber\\   
   &\left.
   \qquad+\frac{1}{\sqrt{X_{12}}} \Big(v_{12}\left(v_{13}+v_{23}-v_{12}\right)-v_{11} v_{23}- v_{22}v_{13}
  +v_{11}v_{22}\Big) \log
   \!\left(\frac{v_{12}+\sqrt{X_{12}}}{v_{12}-\sqrt{X_{12}}}\right)
\right]\nonumber\\
&+\frac{\pi}{2 v_{12} v_{23} v_{13}-v_{11} v_{23}^2-v_{22} v_{13}^2}
\nonumber\\
&\times
\left[
\frac{1}{\sqrt{X_{12}}} \Big(v_{12}\left(v_{13}+v_{23}-v_{12}\right)-v_{11} v_{23}- v_{22}v_{13}
  +v_{11}v_{22}\Big)f_1^{(2)}(v_{12},v_{11},v_{22})
   \right.
   \nonumber\\
&\qquad 
- \Big(v_{13}-v_{12}-v_{23}+\frac{v_{11}
   v_{23}}{v_{13}}\Big) f_1^{(1)}(v_{13},v_{11})
   -
   \Big(v_{23}-v_{12}-v_{13}+\frac{v_{22}v_{13}}{v_{23}} 
   \Big) f_1^{(1)}(v_{23},v_{22})
   \nonumber\\&\left.
\quad
   +2\sqrt{Y_{123}^{(2)}}\,f_Y^{(2)}(v_{12},v_{13},v_{23},v_{11},v_{22})
\right]+\mathcal{O}\!\left(\eps^2\right)\,.
\end{align}
with $f_Y^{(2)}$ from eq.\,\eqref{eq: fY2}.
Finally, using the kinematic quantities from section \ref{sec: Dimensional shift identity}, the triple-massive three-denominator integral has the $\eps$-expansion
\begin{align}
I_{1,1,1}^{(3)}&=\frac{\pi}{X_{123}^{(3)}}\left[\frac{X_{12\bar{3}}}{\sqrt{X_{12}}}\log\!\left(\frac{v_{12}+\sqrt{X_{12}}}{v_{12}-\sqrt{X_{12}}}\right)
+\frac{X_{1\bar{2}3}}{\sqrt{X_{13}}}\log\!\left(\frac{v_{13}+\sqrt{X_{13}}}{v_{13}-\sqrt{X_{13}}}\right)
\right.\nonumber\\
&\left.\qquad\qquad
+\frac{X_{\bar{1}23}}{\sqrt{X_{23}}}\log\!\left(\frac{v_{23}+\sqrt{X_{23}}}{v_{23}-\sqrt{X_{23}}}\right)
\right]\nonumber\\
&+\frac{\pi\,\eps}{X_{123}^{(3)}}\left[\frac{X_{12\bar{3}}}{\sqrt{X_{12}}}f_1^{(2)}(v_{12},v_{11},v_{22})
+\frac{X_{1\bar{2}3}}{\sqrt{X_{13}}}f_1^{(2)}(v_{13},v_{11},v_{33})
+\frac{X_{\bar{1}23}}{\sqrt{X_{23}}}f_1^{(2)}(v_{23},v_{22},v_{33})
\right.\nonumber\\
&\left.\qquad\qquad
+2\sqrt{Y_{123}^{(3)}}\,f_Y^{(3)}
\right]+\mathcal{O}\!\left(\eps^2\right)
\end{align}
with $f_Y^{(3)}$ from eq.\,\eqref{eq: fY3}.
For the double- and triple-massive integral, poles and small-mass limit of the finite part are also in agreement with the results from \cite{Smirnov:2024}.

\section{Beyond three denominators}
\label{sec: Beyond three denominators}
If we consider a situation in $d=4-2\eps$ dimensions where all vectors $v_i$ are confined to the physical $4$-dimensional subspace -- an assumption applicable if the $v_i$ correspond to momenta of observed particles -- we can reduce angular integrals with an arbitrary number of denominators raised to integer powers to the three-denominator case presented in this paper.
This is possible since in $4$ dimensions, i.e. three spatial dimensions, no more than three $\vec{v}_i$ can be linearly dependent.
Hence for any $n>3$ and vectors $v_1,\dots,v_n$ there are constants $\lambda_i$ such that $\lambda\equiv\sum_{i=1}^n \lambda_i\neq 0$ but $0=\sum_{i=1}^n\lambda_i\vec{v}_i$.
By partial fractioning it holds that \cite{Lyubovitskij:2021}
\begin{align}
\prod_{i=1}^n\frac{1}{v_i\cdot k}=\frac{1}{\lambda}\sum_{j=1}^n\lambda_j \prod_{\overset{i=1}{i\neq j}}^n\frac{1}{v_{i}\cdot k}\,,
\end{align}
which reduces the number of denominators in each term.
Iterative application of this identity reduces any $I_{j_1,\dots,j_n}$ with integer-valued $j_i$ to the three-denominator case.
Therefore the presented results can be applied to this much wider class of angular integrals.

\section{Conclusion}
\label{sec: Conclusion}
In this work we performed the first systematic study of angular integrals with three denominators and an arbitrary number of masses.
From IBP relations we derived explicit recursion relations for the reduction to a small set of master integrals.
The $\eps$-expansion of the master integrals was calculated using the method of differential equations.

This methodology, which is well established for loop integrals, was combined with a dimensional-shift identity for angular integrals which allows for an extraction of the pole part before calculating the integral.
In the case of three denominators the order $\eps^0$ also turns out to be fully determined by two-denominator integrals, the unknown genuine three-denominator part only starts contributing at order $\eps$.
The remaining non-trivial part at order $\eps$, which contributes proportional to a Euclidean Gram determinant, has been expressed in terms of Clausen functions.
In the massless case we were able to give a geometrical interpretation of all involved quantities.
The ancillary \texttt{Mathematica} file provides the function \verb!AngularIntegral[{j1_,j2_,j3_},{v12_,v13_,v23_,v11_,v22_,v33_}]! which evaluates the three-denominator angular integral in $d=4-2\eps$ dimensions up to order $\eps$.

In parallel work, Taushif Ahmed, Syed Mehedi Hasan, and Andreas Rapakoulias independently calculated the angular integral with three denominators using Mellin-Barnes integrals \cite{Ahmed:2024}.
Preliminary numerical comparison at selected phase space points showed agreement.
\begin{acknowledgments}
The authors thank François Arleo and Stéphane Munier for organizing the outstanding QCD Masterclass 2023 and 2024 in Saint-Jacut-de-la-Mer, where Johannes Henn's insightful lectures on differential equations helped to inspire this work.
We are grateful to Werner Vogelsang for the possibility to pursue this interesting topic as part of our PhD projects.
We also thank Taushif Ahmed, Syed Mehedi Hasan, and Andreas Rapakoulias for numerical comparison with their independent calculation.
F.W. is indebted to Vladimir A. Smirnov for pleasant discussions on related topics.
This work has been supported by Deutsche Forschungsgemeinschaft (DFG) through the Research Unit FOR 2926 (project 409651613).
\end{acknowledgments}
\appendix
\section{Scalar products for mass reduction formula}
\label{app: Scalar products for mass reduction formula}
The scalar products involving the massless auxiliary vectors appearing in section \ref{sec: Reduction of masses} are given by ($i,j,k\in \lbrace 1,2,3\rbrace$)
\begin{align}
v_{i(ij)}&=\frac{X_{ij}+(v_{ii}-v_{ij})\sqrt{X_{ij}}}{2v_{ij}-v_{ii}-v_{jj}}\,,\\
v_{(ij)(ji)}&=\frac{2 X_{ij}}{2v_{ij}-v_{ii}-v_{jj}}\,,\\
v_{k(ij)}&=\frac{v_{ij}v_{jk}-v_{ik}v_{jj}+v_{ij}v_{ik}-v_{ii}v_{jk}+(v_{ik}-v_{jk})\sqrt{X_{ij}}}{2v_{ij}-v_{ii}-v_{jj}}\,,\\
v_{(ij)(ik)}&=\frac{1}{(2v_{ij}-v_{ii}-v_{jj})(2v_{ik}-v_{ii}-v_{kk})}
\nonumber\\
\times&\Big[
v_{ii}^2 v_{jk}-v_{ii} v_{ij} v_{ik}-v_{ii} v_{ij} v_{jk}-v_{ii} v_{ik} v_{jk}+v_{ii} v_{jj} v_{kk}+v_{ij} v_{ik}^2+v_{ij}^2 v_{ik}-v_{jj}
   v_{ik}^2
\nonumber
\\
&+v_{ij} v_{ik} v_{jk}-v_{kk} v_{ij}^2
   -\sqrt{X_{ij}} \left(v_{ij} v_{ik}+v_{ik} v_{jk}-v_{ii} v_{jk}-v_{kk}
   v_{ij}-v_{ik}^2+v_{ii} v_{kk}\right)
\nonumber
\\
&   
   -\sqrt{X_{ik}} \left(v_{ij} v_{ik}+v_{ij} v_{jk}-v_{jj} v_{ik}-v_{ii} v_{jk}-v_{ij}^2+v_{ii}
   v_{jj}\right)
   \nonumber
\\
&  +\left(-v_{ij}-v_{ik}+v_{ii}+v_{jk}\right) \sqrt{X_{ij}X_{ik}}
\Big]\,,
\\
v_{(ij)(ki)}&=\frac{1}{(2v_{ij}-v_{ii}-v_{jj})(2v_{ik}-v_{ii}-v_{kk})}
\nonumber\\
\times&\Big[
v_{ii}^2 v_{jk}-v_{ii} v_{ij} v_{ik}-v_{ii} v_{ij} v_{jk}-v_{ii} v_{ik} v_{jk}+v_{ii} v_{jj} v_{kk}+v_{ij} v_{ik}^2+v_{ij}^2 v_{ik}-v_{jj}
   v_{ik}^2
\nonumber
\\
&\qquad+v_{ij} v_{ik} v_{jk}-v_{kk} v_{ij}^2
   -\sqrt{X_{ij}} \left(v_{ij} v_{ik}+v_{ik} v_{jk}-v_{ii} v_{jk}-v_{kk}
   v_{ij}-v_{ik}^2+v_{ii} v_{kk}\right)
\nonumber
\\
&\qquad      
   -\sqrt{X_{ik}} \left(-v_{ij} v_{ik}-v_{ij} v_{jk}+v_{jj} v_{ik}+v_{ii} v_{jk}+v_{ij}^2-v_{ii}
   v_{jj}\right)
\nonumber\\&   
   +\left(v_{ij}+v_{ik}-v_{ii}-v_{jk}\right) \sqrt{X_{ij}X_{ik}}
   \Big]\,,
\end{align}
where $X_{ij}=v_{ij}^2-v_{ii}v_{jj}$\,.
\section{Full form of IBP relations}
\label{app: IBP relations}
Plugging the six different $\vec{\xi}$ given in eqs.\,\eqref{eq: Def xi_ik} and \eqref{eq: Def xi_ij} into eq.\eqref{eq: IBP relations product rule} we obtain the following set of IBP relations:
%\begin{widetext}
\begin{align}
&\text{for }\vec{\xi}_{1k}\text{: }
0=\left(-d+j_1+j_2+j_3+2\right) I_{j_1-1,j_2,j_3}
+\left(d-2 j_1-j_2-j_3-2\right) I_{j_1,j_2,j_3}
\nonumber
\\&
\;\;+j_3\,v_{13}\, I_{j_1,j_2,j_3+1}
+j_2\,v_{12}\,I_{j_1,j_2+1,j_3}
+j_1\,v_{11}\,I_{j_1+1,j_2,j_3}
-j_3\,I_{j_1-1,j_2,j_3+1}
-j_2\,I_{j_1-1,j_2+1,j_3} \,,
\\
&\text{for }\vec{\xi}_{12}\text{: }
0=j_3 \left(v_{23}-1\right) I_{j_1-1,j_2,j_3+1}
+j_2 \left(v_{22}-1\right) I_{j_1-1,j_2+1,j_3}
-j_3 \left(v_{13}-1\right) I_{j_1,j_2-1,j_3+1}
\nonumber
\\&
\;\;+\left(j_1-j_2\right)\left(v_{12}-1\right) I_{j_1,j_2,j_3}
+j_3 \left(v_{13}-v_{23}\right) I_{j_1,j_2,j_3+1}
+j_2 \left(v_{12}-v_{22}\right) I_{j_1,j_2+1,j_3}
\nonumber
\\&
\;\;-j_1\left(v_{11}-1\right) I_{j_1+1,j_2-1,j_3}
+j_1 \left(v_{11}-v_{12}\right) I_{j_1+1,j_2,j_3}
\end{align}
%\end{widetext}
IBP relations for \change{$\vec{\xi}_{2k}$, $\vec{\xi}_{3k}$, $\vec{\xi}_{13}$, and $\vec{\xi}_{23}$} follow by appropriate permutation of indices.
\section{Recursion relations for three-denominator angular integrals}
\label{app: Recursion relations}
By taking suitable linear combinations, the IBP relations can be cast into explicit recursions.
A recursion relation that lowers the sum of indices $j=j_1+j_2+j_3$, applicable for $j_1\neq 1$, is given by
%\begin{widetext}
\begin{align}
&I_{j_1,j_2,j_3}=\frac{1}{(1-j_1)X_{123}}\Big[
-\left(v_{23}^2-v_{22} v_{33}\right) \left(-d+j+1\right) I_{j_1-2,j_2,j_3}
\nonumber
\\
& 
+\left(v_{13} v_{23}-v_{12} v_{33}\right)
   \left(-d+j+1\right) I_{j_1-1,j_2-1,j_3}
   -\left(v_{13} v_{22}-v_{12} v_{23}\right) \left(-d+j+1\right)
   I_{j_1-1,j_2,j_3-1}
   \nonumber
\\
& 
+[(v_{12} v_{23}-v_{13} v_{22}) (d-j_1-j_2-2 j_3-1)
+(v_{13} v_{23}-v_{12} v_{33}) (d-j_1-2 j_2-j_3-1)
\nonumber
\\
&\quad+\left(v_{22}
   v_{33}-v_{23}^2\right) (d-2 j_1-j_2-j_3)]I_{j_1-1,j_2,j_3}
\nonumber
\\
&   
   +j_3 \left(v_{23}^2-v_{22} v_{33}\right) I_{j_1-2,j_2,j_3+1}   
   +j_2 \left(v_{23}^2-v_{22} v_{33}\right)
   I_{j_1-2,j_2+1,j_3}
\nonumber
\\
&   
   +j_3 \left(v_{12} v_{33}-v_{13} v_{23}\right) I_{j_1-1,j_2-1,j_3+1}  
   +j_2 \left(v_{13} v_{22}-v  _{12} v_{23}\right)
   I_{j_1-1,j_2+1,j_3-1}
\nonumber
\\
&    
   +\left(j_1-1\right) \left(v_{12} v_{33}-v_{13} v_{23}\right) I_{j_1,j_2-1,j_3}  
   +\left(j_1-1\right) \left(v_{13} v_{22}-v_{12}
   v_{23}\right) I_{j_1,j_2,j_3-1}
\Big] ,
\end{align}
%\end{widetext}
where $X_{123}$ is the Gram determinant defined in section \ref{sec: Dimensional shift identity}.

There are also the symmetric identities with $1\leftrightarrow 2$ and $1\leftrightarrow 3$ applicable for $j_{2,3}\neq 1$.
These identities allow for a systematic reduction of any $I_{j_1,j_2,j_3}$ with $j_1,j_2,j_3>0$ to the case where either $j_1=j_2=j_3=1$ or one of the indices becomes $0$.
From there, recursion relations for two denominators can be used for further reduction \cite{Lyubovitskij:2021}.
To deal with integrals with a negative index, we can use 
%\begin{widetext}
\begin{align}
&I_{j_1,j_2,j_3}=\frac{1}{(3+j-d)Y_{23}}
\Big[
 [v_{23}^2 \left(-d+2 j_1+j_2+j_3+4\right)
\nonumber\\
&\quad
 +v_{23} \left(j_2 \left(v_{12}-3\right)+\left(j_3-1\right) v_{13}-j_1
   \left(v_{12}+v_{13}+2\right)-v_{12}+2 d-3 j_3-6\right)
  \nonumber\\
& \quad 
   +v_{33} \left(\left(j_1-j_2+1\right) v_{12}-d+j_1+2 j_2+j_3+3\right) 
   +v_{22} \left(d
   \left(v_{33}-1\right)+j_1 v_{13}-j_3 v_{13}
\right.
     \nonumber\\
& \quad\left.\quad  -\left(2 j_1+j_2+j_3+4\right) v_{33}+v_{13}+j_1+j_2+2 j_3+3\right)]
I_{j_1+1,j_2,j_3}
\nonumber
\\
&     
   +j_3 \left(v_{13}-1\right)
   \left(v_{23}-v_{33}\right) I_{j_1+1,j_2-1,j_3+1}
 -j_3 \left(v_{13}-1\right) \left(v_{23}^2-v_{22} v_{33}\right) I_{j_1+1,j_2,j_3+1}
\nonumber
\\
&   
-j_2
   \left(v_{12}-1\right) \left(v_{22}-v_{23}\right) I_{j_1+1,j_2+1,j_3-1}-j_2 \left(v_{12}-1\right) \left(v_{23}^2-v_{22} v_{33}\right)
   I_{j_1+1,j_2+1,j_3}
   \nonumber
\\
&  
+\left(j_1+1\right) \left(v_{11}-1\right) \left(v_{23}-v_{33}\right) I_{j_1+2,j_2-1,j_3}-\left(j_1+1\right) \left(v_{11}-1\right)
   \left(v_{22}-v_{23}\right) I_{j_1+2,j_2,j_3-1}
\nonumber
\\
&     
   -\left(j_1+1\right) \left(v_{13} \left(v_{22}-v_{23}\right)+v_{12} \left(v_{33}-v_{23}\right)+v_{11}
   \left(v_{23}^2-v_{22} v_{33}\right)\right) I_{j_1+2,j_2,j_3}
\Big],
\end{align}
%\end{widetext}
where
\begin{align}
Y_{23}=(\vec{v}_2\cdot \vec{v}_3)^2-\vec{v}_2^2\,\vec{v}_3^2=v_{22} + v_{33} - v_{22} v_{33} - v_{23} (2 - v_{23})\,,
\end{align}
which raises $j_1+j_2+j_3$ as well as $j_1$.
Again there are the symmetric versions that raise $j_2$ respectively $j_3$.
These can be used to increase any negative index to $0$ again resulting in two-denominator integrals.
We note that $Y_{23}$ has the geometrical interpretation $Y_{23}=4 \vol^2(\vec{v}_2,\vec{v}_3)$, where $\vol(\vec{v}_2,\vec{v}_3)$ denotes the area of the triangle spanned by $\vec{v}_{2,3}$.

\section{Proof of the general dimensional-shift identity}
\label{app: Proof of the general dimensional shift identity}
In this appendix we prove the general dimensional-shift identity eq.\,\eqref{eq: General dimensional shift identity}.
The proof is a straightforward generalization of the two-denominator case given in \cite{Lyubovitskij:2021}.
We start from the hypergeometric representation of the one-denominator angular integral in $d=4-2\eps$ dimensions,
\begin{align}
I^{(1)}_j(v^2;\eps)=\frac{2\pi}{1-2\eps}\,\ghy\!\left(\frac{j}{2},\frac{j+1}{2},\frac{3}{2}-\eps;1-v^2\right).
\end{align}
Using the contiguous neighbors relation
\begin{align}
\ghy(a,b,c-1;z)=\frac{1}{c-1}\Big[a\,\ghy(a+1,b,c;z)+(c-1-a)\,\ghy(a,b,c;z)\Big] ,
\end{align}
and the symmetry of the hypergeometric function in its first two arguments we can identify the hypergeometric functions with one-denominator integrals with $\eps\rightarrow\eps-1$,
\begin{align}
I^{(1)}_j(v^2;\eps)=\frac{1}{1-2\eps}\left[j\,I^{(1)}_{j+1}(v^2;\eps-1)+(3-j-2\eps)\,I^{(1)}_j(v^2;\eps-1)\right]\,.
\label{eq: One-denominator dim shift}
\end{align} 
Now considering the general case of $n$ denominators, we use a Feynman parametrization of the denominator and obtain
\begin{align}
I_{j_1,\dots,j_n}&=\int\frac{\dx\Omega_{d-1}}{\Omega_{d-3}}\frac{1}{(v_1\cdot k)^{j_1}\cdots (v_n\cdot k)^{j_n}}\,\nonumber\\
&=\frac{\Gamma(j)}{\Gamma(j_1)\cdots\Gamma(j_n)}\int_0^1\dx x_1\cdots\int_0^1\dx x_n\,x_1^{j_1}\cdots x_n^{j_n}\int\frac{\dx\Omega_{d-1}}{\Omega_{d-3}}\frac{\delta\!\left(1-\sum_{i=1}^n x_i\right)}{\left(\sum_{i=1}^nx_i v_i\cdot k\right)^j}
\nonumber\\
&=\frac{\Gamma(j)}{\Gamma(j_1)\cdots\Gamma(j_n)}\int_0^1\dx x_1\cdots\int_0^1\dx x_n\,x_1^{j_1}\cdots x_n^{j_n}\,\delta\!\left(1-\sum_{i=1}^n x_i\right) I_{j}^{(1)}\!\left(\left(\sum_{i=1}^nx_i v_i\right)^2;\eps\right),
\end{align}
where $j=\sum_{i=1}^n j_i$.
Plugging in eq.\,\eqref{eq: One-denominator dim shift}, the second summand can be directly identified with an $n$-denominator angular integral with $\eps\rightarrow\eps-1$.
For the first summand, realizing that
\begin{align}
x_1^{j_1}\cdots x_n^{j_n}\,\delta\!\left(1-\sum_{i=1}^n x_i\right)=\left[x_1^{j_1+1}x_2^{j_2}\cdots x_n^{j_n}+\cdots+x_1^{j_1}x_2^{j_2}\cdots x_n^{j_n+1}\right]\delta\!\left(1-\sum_{i=1}^n x_i\right) ,
\end{align}
we achieve $j\rightarrow j+1$ in the Feynman parameters and can thus identify each term with an $n$-denominator angular integral with $\eps\rightarrow\eps-1$ resulting in
\begin{align}
I_{j_ 1,...,j_n}(\eps)=&\frac{1}{1-2\eps}\left[\sum_{i=1}^n j_i I_{j_1,...,j_i+1,...,j_n}(\eps-1)
%\right.
%\nonumber\\
%&\left.
+\,\left(3-\sum_{i=1}^n j_i-2\eps\right)I_{j_1,...,j_n}(\eps-1)\right],
\end{align}
which is eq.\,\eqref{eq: General dimensional shift identity} from the main text.

\section{Expansion of dimensionally-shifted two-denominator integrals}
\label{app: Expansion of dimensionally shifted two-denominator integrals}
In this appendix we list the $\eps$-expansions of the dimensionally-shifted two-denominator integrals used in the differential equation in section \ref{sec: Integration for master integrals}.
They are constructed from the known expansion in $d=4-2\eps$ dimensions \cite{Beenakker:1988, Somogyi:2011, Lyubovitskij:2021} via the identities
\begin{align}
I^{(0)}(\eps-1)&=\frac{1-2\eps}{3-2\eps}\,I^{(0)}(\eps)\,,\\
I^{(1)}_1(v_{11};\eps-1)&=\frac{1-2\eps}{2(1-\eps)(1-v_{11})}\Big[I^{(0)}(\eps)-v_{11}\,I^{(1)}_1(v_{11};\eps)\Big]\,,\\
I^{(0)}_{1,1}(v_{12};\eps-1)&=-\frac{1}{2-v_{12}}\left[\frac{1-2\eps}{\eps}\,I^{(0)}(\eps)+v_{12}\,I^{(0)}_{1,1}(v_{12};\eps)\right],\\
I^{(1)}_{1,1}(v_{12},v_{11};\eps-1)&=-\frac{v_{12}}{v_{12}(2-v_{12})-v_{11}}\left[\frac{1-2\eps}{2\eps}\,I^{(0)}(\eps)-\left(1-\frac{v_{11}}{v_{12}}\right)I^{(1)}_1(v_{11};\eps)
\right.\\
\nonumber
&\left.
\qquad\qquad\qquad\qquad+v_{12}\,I^{(1)}_{1,1}(v_{12},v_{11};\eps)\right],
\end{align}
that follow from applying eq.\,\eqref{eq: General dimensional shift identity} in the cases of zero, one, and two denominators in combination with IBP reduction to master integrals and subsequently solving for the integrals in $d=4-2\eps$ dimensions.

For the massless two-denominator integral, required in the integration of the massless three-denominator integral, we find the expansion
\begin{align}
I_{1,1}^{(0)}(v_{12},\eps-1)=-\frac{2\pi\log\left(\frac{v_{12}}{2}\right)}{2-v_{12}}+\mathcal{O}(\eps)\,.
\label{eq: I11 massless d=6}
\end{align}
For the massive integral we further need the expansions
\begin{align}
 I^{(0)}(\eps-1)&=\frac{2\pi}{3}+\mathcal{O}(\eps)\,,\\
    I_{2}^{(1)}(v_{11};\eps-1)&=-\frac{2 \pi }{1-v_{11}}+\frac{\pi  \log\!
   \left(\frac{1+\sqrt{1-v_{11}}}{1-\sqrt{1-v_{11}}}\right)}{\left(1-v_{11}\right){}^{3/2}}+\mathcal{O}(\eps)\,,
   \\
I_{1,1}^{(1)}(v_{12},v_{11};\eps-1)&=\frac{\pi  \left(v_{12}-v_{11}\right) \log
   \!\left(\frac{1+\sqrt{1-v_{11}}}{1-\sqrt{1-v_{11}}}\right)}{\sqrt{1-v_{11}}
   \left(v_{12}\left(2-v_{12}\right)-v_{11}\right)}+\frac{\pi\, v_{12} \log
  \frac{v_{11}}{v_{12}^2}}{v_{12}\left(2-v_{12}\right)-v_{11}
  }+\mathcal{O}(\eps)\,.
\label{eq: I11 massive d=6}
\end{align}
\section{Calculation of the boundary value for the massless master integral}
\label{app: Calculation of the boundary value for the massless master integral}
This appendix is dedicated to the calculation of the boundary value used in the massless master integral.
The approach is a direct calculation using a string of series, integral, and special function identities successively reducing the complexity of the original integral.
This will finally lead us to a remarkably compact result.

The boundary value for $J^{(0),d=6}_{1,1,1}$ is calculated at the symmetric point $v_{12}=v_{13}=v_{23}=1$, where the three vectors are in an orthogonal configuration depicted in figure \ref{fig:Boundary value}.
At this point the Euclidean Gram determinant becomes unity and the defining integral representation eq.\,\eqref{eq: General angular integral} becomes
\begin{align}
&J^{(0),d=6}_{1,1,1}(1,1,1)=I^{(0),d=6}_{1,1,1}(1,1,1)
\nonumber\\
&=\frac{1}{2}\int_0^\pi\dx\theta_1\int_0^\pi\dx\theta_2\int_0^\pi\dx\theta_3\frac{\sin^3\theta_1\sin^2\theta_2\sin\theta_3}{(1-\cos\theta)(1-\sin\theta_1\cos\theta_2)(1-\sin\theta_1\sin\theta_2\cos\theta_3)}\,.
\label{eq: J111 boundary value}
\end{align}
Expanding the third denominator in a geometric series and integrating the $\theta_3$-integrals\footnote{\change{To be mathematically precise, care must be taken whenever interchanging a series with an integration.
In this particular case one can convince oneself that the partial sums of the geometric series are -- for an odd upper bound -- bounded by the non-expanded expression, so it only remains to show that the original integral is finite for Lebesgue's dominated convergence theorem to be applicable.
Finiteness of the original integral can be established by checking that the integrand is finite at the spurious singularities at $\theta_1=0$ respectively $\theta_1=\pi/2$, $\theta_2=0$, and only grows as $r^{-1}$ ($r$ being the distance to the singularity) near $(\theta_1,\theta_2,\theta_3)=(\pi/2,\pi/2,0)$ which is integrable in a $3$ dimensional integral.\label{foot:Lebesgue}}}
, using that
\begin{align}
\int_0^\pi\dx\theta\sin\theta \cos^n\theta=\frac{1+(-1)^n}{n+1}
\end{align}
vanishes for odd $n$, we get to
\begin{align}
J^{(0),d=6}_{1,1,1}(1,1,1)&=\sum_{n=0}^\infty\frac{1}{2n+1}\int_0^\pi\dx\theta_1\int_0^\pi\dx\theta_2\frac{\sin^{3+2n}\theta_1\sin^{2+2n}\theta_2}{(1-\cos\theta)(1-\sin\theta_1\cos\theta_2)}
\nonumber\\
&=\sum_{n=0}^\infty\frac{1}{2n+1}\,I_{1,1}^{(0)}(v_{12}=1;\eps=-n-1)\,,
\label{eq: boundary value sum 1}
\end{align}
where we could identify the remaining integral with a massless two-denominator angular integral in $6+2n$ dimensions.
This integral has been calculated in $d$ dimensions in terms of the Gauss hypergeometric function \cite{vanNeerven:1985,Somogyi:2011,Lyubovitskij:2021},
\begin{align}
I_{1,1}^{(0)}(v_{12};\eps)=-\frac{\pi}{\eps}\,\ghy\!\left(1,1,1-\eps;1-\frac{v_{12}}{2}\right).
\end{align}
In the specific case at hand we can further use the identity
\begin{align}
\ghy\!\left(1,1,c;\frac{1}{2}\right)=(c-1)\left[\digam\!\left(\frac{c}{2}\right)-\digam\!\left(\frac{c}{2}-\frac{1}{2}\right)\right],
\end{align}
reducing the hypergeometric function to digamma functions $\digam(z)\equiv\frac{\dx}{\dx z}\Gamma(z)$.
This allows for the representation
\begin{align}
I_{1,1}^{(0)}\!\left(\change{1};-n-1\right)=\frac{\pi}{n+1}\,\ghy\left(1,1,2+n;\frac{1}{2}\right)=\pi\left[\digam\!\left(\frac{n}{2}+1\right)-\digam\!\left(\frac{n}{2}+\frac{1}{2}\right)\right].
\end{align}
Putting this into eq.\,\eqref{eq: boundary value sum 1} and splitting the sum in even and odd terms we have
\begin{align}
J^{(0),d=6}_{1,1,1}(1,1,1)&=\pi\,\sum_{n=0}^\infty \frac{1}{4n+1}\left[\digam\!\left(n+1\right)-\digam\!\left(n+\frac{1}{2}\right)\right]
\nonumber\\&
+\pi\,\sum_{n=0}^\infty \frac{1}{4n+3}\left[\digam\!\left(n+\frac{3}{2}\right)-\digam\!\left(n+1\right)\right].
\end{align}
The digamma function at integer respectively half-integer values evaluates to
\begin{align}
\digam\!\left(n\right)&=H_{n-1}-\gammaeuler\,,\\
\digam\!\left(n+\frac{1}{2}\right)&=2\, U_n-\gammaeuler-2\log 2\,,\\
\end{align}
where $H_n\equiv\sum_{k=1}^n 1/k$ denotes the harmonic numbers and $U_n=\sum_{k=1}^n 1/(2k-1)$ the odd harmonic numbers.
Using the identity
\begin{align}
H_n-2\,U_n=-2 A_{2n}
\end{align}
where $A_n=\sum_{k=1}^n (-1)^{k+1}/k$ denotes the alternating harmonic \change{numbers} we find
\begin{align}
\digam\!\left(n+1\right)-\digam\!\left(n+\frac{1}{2}\right)&=-2 A_{2n}+2\log 2\,,\\
\digam\!\left(n+\frac{3}{2}\right)-\digam\!\left(n+1\right)&=2 A_{2n+2}-2\log 2\,.
\end{align}
Combining even and odd terms from both sums we have the representation
\begin{align}
J^{(0),d=6}_{1,1,1}(1,1,1)=2\pi\sum_{n=0}^\infty\frac{(-1)^n}{2n+1}\left(\log 2-A_n\right).
\label{eq: J111 boundary An representation}
\end{align}
Starting from the integral representation
\begin{align}
\log 2-A_n=\int_0^1\dx x\,\frac{(-x)^n}{1+x}\,,
\end{align}
substituting $x=\left(\frac{1-t}{1+t}\right)^2$, and performing partial integration with respect to $1/(1+t^2)$ leads to the representation
\begin{align}
\log 2-A_n=4(-1)^n(2n+1)\int_0^1\frac{\dx t}{(1+t)^2}\left(\frac{1-t}{1+t}\right)^{2n}\,\arctan t\,.
\label{eq: An integral representation}
\end{align}
Plugging this into eq.\,\eqref{eq: J111 boundary An representation} and summing the geometric series we finally obtain
\begin{align}
J^{(0),d=6}_{1,1,1}(1,1,1)=2\pi\,\int_0^1\frac{\dx t}{t}\,\arctan t=2\pi\,\mathrm{Ti}_2(1)=2\pi\sum_{n=1}^\infty \frac{(-1)^{n+1}}{(2n+1)^2}=2\pi\,\catalan\,,
\end{align}
where $\mathrm{Ti}_2$ is the inverse tangent integral and $\catalan$ is Catalan's constant.
\change{A direct numerical check of the original integral eq.\,\eqref{eq: J111 boundary value} additionally confirms $J_{1,1,1}^{(0),d=6}(1,1,1)\approx 5.75518$ further raising confidence in the correctness of the result also for physicists who are not inclined to go through the motions akin to footnote \ref{foot:Lebesgue} in each step.}

\section{Useful integrals for the calculation of the master integrals}
\label{app: Useful integrals for the calculation of the master integrals}
From Lewin's delightful book \cite{Lewin:1981} we have the following integrals useful for the integration of eqs. \eqref{eq: Integration kernel massless order eps} and \eqref{eq: Integration kernel massive order eps},
\begin{align}
J_0(\delta,x)&=\int_0^x\dx t\,\frac{\delta}{\delta^2+t^2}=\arctan\left(\frac{x}{\delta}\right)\,,
\\
J_1(\alpha,\delta,x)&=\int_{-\alpha}^x\dx t\,\frac{\delta\log|\alpha+t|}{\delta^2+t^2}=\frac{\theta+\eta}{2}\log\left(\alpha^2+\delta^2\right)-\frac{1}{2}\clausen(2\theta+2\eta)
\nonumber\\&\qquad\qquad\qquad\qquad\qquad-\frac{1}{2}\clausen(\pi-2\theta)-\frac{1}{2}\clausen(\pi-2\eta)\,,
\\
J_2(\beta,\delta,x)&=\int_0^x\dx t\,\frac{\delta \log(\beta^2+t^2)}{\delta^2+t^2}
%\nonumber\\
%&
=-\varphi \log\left| \frac{\beta -\delta }{\beta +\delta }\right| 
+\theta \log|
   (\beta -\delta ) (\beta +\delta )|
   \nonumber\\
&\hphantom{=\int_0^x\dx t\,\frac{\delta \log(\beta^2+t^2)}{\delta^2+t^2}=}
+\frac{1}{2} \clausen\!\left(2 \theta-2 \varphi\right)
+\frac{1}{2} \clausen\!\left(2 \theta+2 \varphi\right)
-\clausen\!\left(2 \theta\right) ,
\end{align}
with the abbreviations
\begin{align}
\theta=\arctan\frac{x}{\delta}\,,
\quad
\eta=\arctan\frac{\alpha}{\delta}\,,
\quad
\varphi=\arctan\frac{x}{\beta}\,.
\end{align}
Here $\clausen$ denotes the Clausen function recalled in appendix \ref{app: Clausen function}.

To simplify the arguments of the Clausen functions in these expressions one may use the addition theorem of the arcus tangent
	\begin{align}
		\arctan x+\arctan y=\arctan\frac{x+y}{1- x y}+\pi\,\mathrm{sgn}(x+y)\,\Theta(x y-1) \,,
		\label{eq: Arctan addition theorem}
	\end{align}
where the $\pi$ term can always be dropped because of the $2\pi$-periodicity of $\clausen$.
\section{Clausen function}
\label{app: Clausen function}
\begin{figure}
\centering
\includegraphics[width=0.45\textwidth]{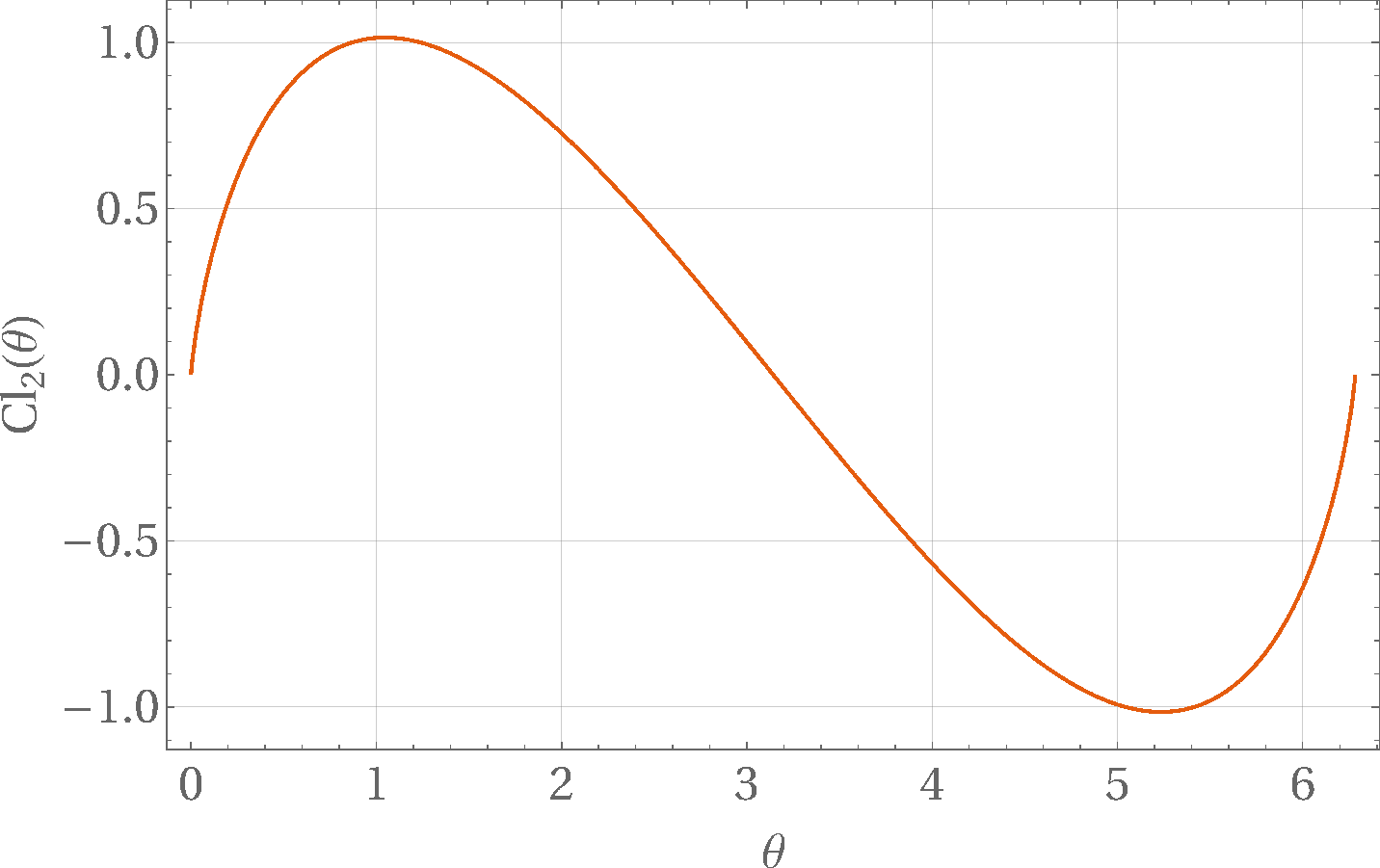}
\hspace{0.5cm}
\includegraphics[width=0.45\textwidth]{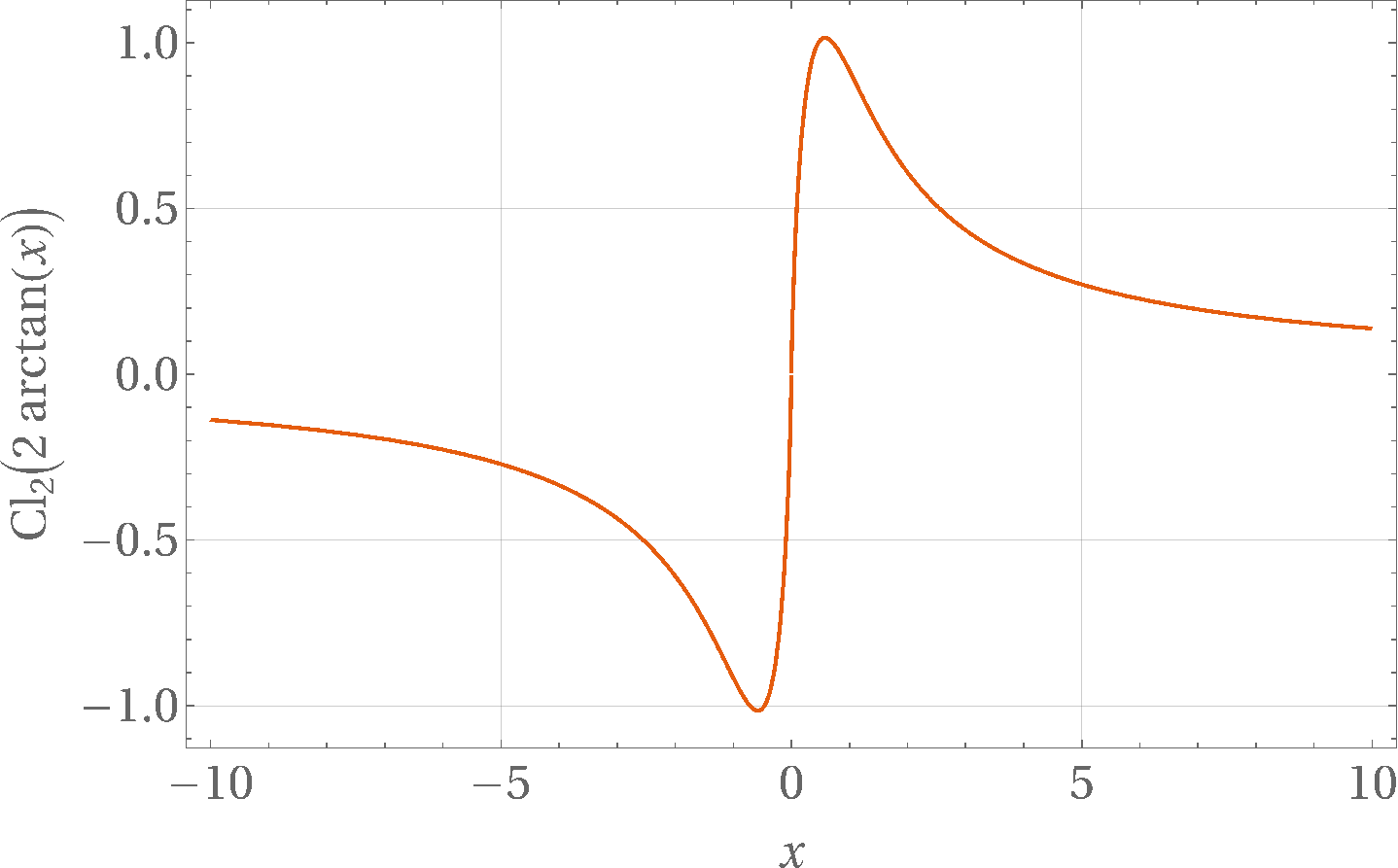}
\caption{The left panel shows a plot of the Clausen function on the interval $[0,2\pi]$. The right panel displays the combination $\clausen(2\arctan x)$ as it appears in the results from section \ref{sec: Results}.}
\label{fig: Clausen function}
\end{figure}

The Clausen function $\clausen(\theta)$, named after 19th century Danish astronomer and mathematician Thomas Clausen, who calculated the function to 16 decimal places in 1832 \cite{Clausen:1832}, is defined by the Fourier series
\begin{align}
\clausen(\theta)=\sum_{n=1}^\infty\frac{\sin(n\theta)}{n^2}\,.
\label{eq: Clausen function Fourier Series}
\end{align}
It arises naturally by considering the dilogarithm of imaginary exponential argument,
\begin{align}
\dilog\!\left(\e^{\iu\theta}\right)=\sum_{n=1}^\infty \frac{\e^{\iu n\theta}}{n^2}=\sum_{n=1}^\infty \frac{\cos n\theta}{n^2}+\iu \sum_{n=1}^\infty \frac{\sin n\theta}{n^2}.
\end{align}
By termwise differentiating the cosine series, one finds the real part to be a polynomial, explicitly for $0\leq \theta\leq 2\pi$,
\begin{align}
\mathrm{Re}\,\dilog\!\left(\e^{\iu\theta}\right)=\frac{\pi^2}{6}-\frac{\theta(2\pi-\theta)}{4}\,,
\end{align}
while the imaginary part is the non-elementary Clausen function,
\begin{align}
\mathrm{Im}\,\dilog\!\left(\e^{\iu\theta}\right)=\clausen(\theta).
\end{align}
By differentiating eq.\,\eqref{eq: Clausen function Fourier Series} one obtains the integral representation \cite{Clausen:1832,Lewin:1981}
\begin{align}
\clausen(\theta)=-\int_0^\theta\dx t\log\!\left|2\sin\frac{t}{2}\right|.
\label{eq: Clausen integral}
\end{align}
In figure \ref{fig: Clausen function} we depict the function $\clausen(\theta)$ as well as the combination $\clausen(2\arctan x)$ appearing in the main text.

The Clausen function is antisymmetric, periodic with period $2\pi$, and obeys $\clausen(\pi+\theta)=-\clausen(\pi-\theta)$.
Furthermore it satisfies the duplication formula
\begin{align}
\clausen(2\theta)=2\,\clausen(\theta)-2\,\clausen(\pi-\theta)\,.
\end{align}
Special values include $\clausen(0)=\clausen(\pi)=0$ and $\clausen(\pi/2)=\catalan$, where $\catalan=\sum_{n=0}^\infty\frac{(-1)^n}{(2n+1)^2}\approx 0.915965594$ denotes Catalan's constant. 
The Clausen function attains its maximum at $\pi/3$ at the value $\clausen(\pi/3)=\gieseking\approx 1.014941606$ called Gieseking's constant.

Using the product expansion $\frac{\sin \pi z}{\pi z}=\prod_{n=1}^\infty \left(1-\frac{z^2}{n^2}\right)$ on the integrand of eq.~\eqref{eq: Clausen integral}, one obtains the series representation
\begin{align}
\clausen(\theta)=\theta-\theta\log|\theta|-\sum_{n=1}^\infty\frac{(-1)^{n}B_{2n}}{2n\,(2n+1)}\frac{\theta^{2n+1}}{(2n)!}\,,
\end{align}
valid for $|\theta|<2\pi$, where $B_{2n}$ are the Bernoulli  numbers.
By keeping one logarithm explicit in the expansion, an even more rapidly convergent series is obtained,
\begin{align}
\clausen(\theta)=3\theta-\theta\log\!\left|\,\theta\left(1-\frac{\theta}{2\pi}\right)\left(1+\frac{\theta}{2\pi}\right)\right|-2\pi\log\!\left(\frac{2\pi+\theta}{2\pi-\theta}\right)+\theta\sum_{n=1}^\infty\frac{\zeta_{2n}-1}{n(2n+1)}\left(\frac{\theta}{2\pi}\right)^{2n}\,,
\end{align}
which utilizes the quick convergence of the even zeta values $\zeta_{2n}\equiv \sum_{k=1}^\infty\frac{1}{k^{2n}}=(-1)^{n-1}\frac{(2\pi)^{2n}B_{2n}}{2 (2n)!}$ to $1$ for large $n$.
It was this series, truncated at $n=14$, Clausen used in his original calculation \cite{Clausen:1832}.
Other efficient numerical implementations can be found in \cite{Wood:1968, Kolbig:1995, Wu:2010}.

The Clausen functions appearing in this paper all have arguments of the form $2\arctan x$ or $2\arctan x+2\arctan y$ -- note that the second type can be brought into the form of the first by using the addition theorem of the arcus tangent \eqref{eq: Arctan addition theorem}. 
For real $x$, $y$ , we can express these Clausen functions in terms of the single-valued Bloch-Wigner dilogarithm $\mathrm{D}_2(z)=\mathrm{Im}\,\dilog(z)+\arg(1-z)\log|z|$ by
\begin{align}
\clausen(2\arctan x)=\mathrm{D}_2\!\left(\frac{\iu-x}{\iu+x}\right)\;\;\text{respectively}\;\;\clausen(2\arctan x+2\arctan y)=\mathrm{D}_2\!\left(\frac{\iu-x}{\iu+x}\,\frac{\iu-y}{\iu+y}\right).
\label{eq: Clausen function as Bloch-Wigner dilogarithms}
\end{align}
The Bloch-Wigner dilogarithm is connected to the volume of ideal hyperbolic tetrahedra \cite{Zagier:2007}.
Concretely, in the upper-half space model, where the hyperbolic space $\frak{H}_3$ is represented by $\mathbb{C}\times\mathbb{R}_+$, a tetrahedron with vertices $z_{0,1,2,3}\in\mathbb{C}$ has the hyperbolic volume 
\begin{align}
\vol_\mathfrak{H}(z_0,z_1,z_2,z_3)=\mathrm{D}_2\!\left(\frac{z_0-z_2}{z_0-z_3}\,\frac{z_1-z_3}{z_1-z_2}\right).
\end{align}
The arguments of the Bloch-Wigner dilogarithms in eq.\,\eqref{eq: Clausen function as Bloch-Wigner dilogarithms} are naturally represented as such cross ratios.
In particular 
\begin{align}
\clausen(2\arctan x+2\arctan y)=\vol_\mathfrak{H}(\iu,-\iu,x,-y)\,.
\end{align}

Another special function that is connected to the Clausen function is the inverse tangent integral $\mathrm{Ti}_2(x)=\int_0^x\frac{\dx t}{t}\,\arctan t$, which can be expressed as
\begin{align}
\mathrm{Ti}_2(x)=\arctan x\,\log x+\frac{1}{2}\clausen(2\arctan x)+\frac{1}{2}\clausen(\pi-2\arctan x)\,.
\end{align}
An analytic continuation of the Clausen function concatenated  with an inverse tangent to complex arguments is given by
\begin{align}
\clausen(2\arctan z)&=
\frac{1}{2 \iu}
\left[\dilog\!\left(\frac{\iu-z}{\iu+z}\right)-\dilog\!\left(\frac{\iu+z}{\iu-z}\right)\right]
\nonumber\\
&=\iu\left(\dilog\!\left(\frac{\iu+z}{\iu-z}\right)+\frac{1}{4}\log^2\!\left(\frac{z+\iu}{z-\iu}\right)+\frac{\pi^2}{12}\right).
\end{align}
It has a branch cut along the imaginary axis from $-\iu$ to $\iu$.
In the high energy physics literature the Clausen function and its related constants appear in a variety of Feynman integral calculations, for example in \cite{Lu:1992,Wagner:1996,Davydychev:1997, Broadhurst:1998,Broadhurst:1999, Davydychev:2000, Davydychev:2003, Czakon:2004, Binger:2006, Adams:2015, Chicherin:2020,Abreu:2022,Devoto:2024}.

\bibliography{Angular_integrals_with_three_denominators}

\providecommand{\href}[2]{#2}\begingroup\raggedright\begin{thebibliography}{10}

\bibitem{Schellekens:1981}
A.~Schellekens, \emph{{Perturbative QCD and lepton pair production}}, Ph.D.
  thesis, Nijmegen University, 6, 1981.

\bibitem{vanNeerven:1985}
W.~van Neerven, \emph{{Dimensional Regularization of Mass and Infrared
  Singularities in Two Loop On-shell Vertex Functions}},
  \href{https://doi.org/10.1016/0550-3213(86)90165-3}{\emph{Nucl. Phys. B}
  {\bfseries 268} (1986) 453}.

\bibitem{Beenakker:1988}
W.~Beenakker, H.~Kuijf, W.~van Neerven and J.~Smith, \emph{{QCD Corrections to
  Heavy Quark Production in p anti-p Collisions}},
  \href{https://doi.org/10.1103/PhysRevD.40.54}{\emph{Phys. Rev. D} {\bfseries
  40} (1989) 54}.

\bibitem{Somogyi:2011}
G.~Somogyi, \emph{{Angular integrals in d dimensions}},
  \href{https://doi.org/10.1063/1.3615515}{\emph{J. Math. Phys.} {\bfseries 52}
  (2011) 083501} [\href{https://arxiv.org/abs/1101.3557}{{\ttfamily
  1101.3557}}].

\bibitem{Lyubovitskij:2021}
{V.\,E. Lyubovitskij, F. Wunder and A.\,S. Zhevlakov}, \emph{{New ideas for
  handling of loop and angular integrals in D-dimensions in QCD}},
  \href{https://doi.org/10.1007/JHEP06(2021)066}{\emph{JHEP} {\bfseries 06}
  (2021) 066} [\href{https://arxiv.org/abs/2102.08943}{{\ttfamily
  2102.08943}}].

\bibitem{Wunder:2024}
F.~Wunder, \emph{Asymptotic behavior of angular integrals in the massless
  limit}, \href{https://doi.org/10.1103/PhysRevD.109.076022}{\emph{Phys. Rev.
  D} {\bfseries 109} (2024) 076022}.

\bibitem{Smirnov:2024}
V.A.~Smirnov and F.~Wunder, \emph{{Expansion by regions meets angular
  integrals}}, \href{https://doi.org/10.1007/JHEP08(2024)138}{\emph{JHEP}
  {\bfseries 08} (2024) 138}
  [\href{https://arxiv.org/abs/2405.13120}{{\ttfamily 2405.13120}}].

\bibitem{Bolzoni:2010}
P.~Bolzoni, G.~Somogyi and Z.~Trocsanyi, \emph{{A subtraction scheme for
  computing QCD jet cross sections at NNLO: integrating the iterated
  singly-unresolved subtraction terms}},
  \href{https://doi.org/10.1007/JHEP01(2011)059}{\emph{JHEP} {\bfseries 01}
  (2011) 059} [\href{https://arxiv.org/abs/1011.1909}{{\ttfamily 1011.1909}}].

\bibitem{Anastasiou:2013}
C.~Anastasiou, C.~Duhr, F.~Dulat and B.~Mistlberger, \emph{{Soft triple-real
  radiation for Higgs production at N3LO}},
  \href{https://doi.org/10.1007/JHEP07(2013)003}{\emph{JHEP} {\bfseries 07}
  (2013) 003} [\href{https://arxiv.org/abs/1302.4379}{{\ttfamily 1302.4379}}].

\bibitem{Lillard:2016}
B.~Lillard, T.M.P.~Tait and P.~Tanedo, \emph{{Kaluza-Klein gluons at 100 TeV:
  NLO corrections}},
  \href{https://doi.org/10.1103/PhysRevD.94.054012}{\emph{Phys. Rev. D}
  {\bfseries 94} (2016) 054012}
  [\href{https://arxiv.org/abs/1602.08622}{{\ttfamily 1602.08622}}].

\bibitem{Kotlarski:2016}
W.~Kotlarski, \emph{{Sgluons in the same-sign lepton searches}},
  \href{https://doi.org/10.1007/JHEP02(2017)027}{\emph{JHEP} {\bfseries 02}
  (2017) 027} [\href{https://arxiv.org/abs/1608.00915}{{\ttfamily
  1608.00915}}].

\bibitem{Lionetti:2018}
S.~Lionetti, \emph{{Subtraction of Infrared Singularities at Higher Orders in
  QCD}}, Ph.D. thesis, {ETH Z\"urich}, 2018.
\newblock 10.3929/ethz-b-000332748.

\bibitem{Specchia:2018}
C.~Specchia, \emph{{Perturbative Corrections to Inclusive and Differential
  Cross Sections for Higgs Production at the LHC}}, Ph.D. thesis, {ETH
  Z\"urich}, 2018.
\newblock 10.3929/ethz-b-000307851.

\bibitem{Bahjat-Abbas:2018}
N.~Bahjat-Abbas, J.~Sinninghe~Damst\'e, L.~Vernazza and C.~White, \emph{{On
  next-to-leading power threshold corrections in Drell-Yan production at
  N$^3$LO}}, \href{https://doi.org/10.1007/JHEP10(2018)144}{\emph{JHEP}
  {\bfseries 10} (2018) 144}
  [\href{https://arxiv.org/abs/1807.09246}{{\ttfamily 1807.09246}}].

\bibitem{Baranowski:2020}
D.~Baranowski, \emph{{NNLO zero-jettiness beam and soft functions to higher
  orders in the dimensional-regularization parameter $\epsilon$}},
  \href{https://doi.org/10.1140/epjc/s10052-020-8047-y}{\emph{Eur. Phys. J. C}
  {\bfseries 80} (2020) 523}
  [\href{https://arxiv.org/abs/2004.03285}{{\ttfamily 2004.03285}}].

\bibitem{Blumlein:2020}
J.~Bl\"umlein, A.~De~Freitas, C.~Raab and K.~Sch\"onwald, \emph{{The
  $O(\alpha^2)$ initial state QED corrections to $e^+e^- \rightarrow
  \gamma^*/Z_0^*$}},
  \href{https://doi.org/10.1016/j.nuclphysb.2020.115055}{\emph{Nucl. Phys. B}
  {\bfseries 956} (2020) 115055}
  [\href{https://arxiv.org/abs/2003.14289}{{\ttfamily 2003.14289}}].

\bibitem{Isidori:2020}
G.~Isidori, S.~Nabeebaccus and R.~Zwicky, \emph{{QED corrections in
  $\overline{B}\to \overline{K}{\mathrm{\ell}}^{+}{\mathrm{\ell}}^{-}$ at the
  double-differential level}},
  \href{https://doi.org/10.1007/JHEP12(2020)104}{\emph{JHEP} {\bfseries 2020}
  (2020) } [\href{https://arxiv.org/abs/2009.00929}{{\ttfamily 2009.00929}}].

\bibitem{Alioli:2022}
S.~Alioli, A.~Broggio and M.A.~Lim, \emph{{Zero-jettiness resummation for
  top-quark pair production at the LHC}},
  \href{https://doi.org/10.1007/JHEP01(2022)066}{\emph{JHEP} {\bfseries 01}
  (2022) 066} [\href{https://arxiv.org/abs/2111.03632}{{\ttfamily
  2111.03632}}].

\bibitem{Assi:2023}
B.~Assi and S.~H\"oche, \emph{{New approach to QCD final-state evolution in
  processes with massive partons}},
  \href{https://doi.org/10.1103/PhysRevD.109.114008}{\emph{Phys. Rev. D}
  {\bfseries 109} (2024) 114008}
  [\href{https://arxiv.org/abs/2307.00728}{{\ttfamily 2307.00728}}].

\bibitem{Catani:2023}
S.~Catani and P.K.~Dhani, \emph{{Collinear functions for QCD resummations}},
  \href{https://doi.org/10.1007/JHEP03(2023)200}{\emph{JHEP} {\bfseries 2023}
  (2023) 1} [\href{https://arxiv.org/abs/2208.05840}{{\ttfamily 2208.05840}}].

\bibitem{Pal:2023}
S.~Pal and S.~Seth, \emph{{On $\mathrm{Higgs}+\mathrm{jet}$ production at
  next-to-leading power accuracy}},
  \href{https://doi.org/10.1103/PhysRevD.109.114018}{\emph{Phys. Rev. D}
  {\bfseries 109} (2024) 114018}
  [\href{https://arxiv.org/abs/2309.08343}{{\ttfamily 2309.08343}}].

\bibitem{Devoto:2024}
F.~Devoto, K.~Melnikov, R.~R\"ontsch, C.~Signorile-Signorile and
  D.M.~Tagliabue, \emph{{A fresh look at the nested soft-collinear subtraction
  scheme: NNLO QCD corrections to N-gluon final states in $ q\overline{q} $
  annihilation}}, \href{https://doi.org/10.1007/JHEP02(2024)016}{\emph{JHEP}
  {\bfseries 02} (2024) 016}
  [\href{https://arxiv.org/abs/2310.17598}{{\ttfamily 2310.17598}}].

\bibitem{Rowe:2024}
M.~Rowe and R.~Zwicky, \emph{{Structure-dependent QED in $B^- \to \ell^- \bar
  \nu (\gamma)$}}, \href{https://doi.org/10.1007/JHEP07(2024)249}{\emph{JHEP}
  {\bfseries 07} (2024) 249}
  [\href{https://arxiv.org/abs/2404.07648}{{\ttfamily 2404.07648}}].

\bibitem{Matsuura:1989}
T.~Matsuura, S.~Van Der~Marck and W.~Van~Neerven, \emph{{The calculation of the
  second order soft and virtual contributions to the Drell-Yan cross section}},
  \href{https://doi.org/https://doi.org/10.1016/0550-3213(89)90620-2}{\emph{Nucl.
  Phys. B} {\bfseries 319} (1989) 570}.

\bibitem{Matsuura:1990}
T.~Matsuura, R.~Hamberg and W.L.~van Neerven, \emph{{The contribution of the
  gluon-gluon subprocess to the Drell-Yan K-factor}},
  \href{https://doi.org/https://doi.org/10.1016/0550-3213(90)90391-P}{\emph{Nucl.
  Phys. B} {\bfseries 345} (1990) 331}.

\bibitem{Hamberg:1991}
R.~Hamberg, W.~Van~Neerven and T.~Matsuura, \emph{{A complete calculation of
  the order $\alpha_s^2$ correction to the Drell-Yan K-factor}},
  \href{https://doi.org/https://doi.org/10.1016/0550-3213(91)90064-5}{\emph{Nucl.
  Phys. B} {\bfseries 359} (1991) 343}.

\bibitem{Mirkes:1992}
E.~Mirkes, \emph{{Angular decay distribution of leptons from W bosons at NLO in
  hadronic collisions}},
  \href{https://doi.org/10.1016/0550-3213(92)90046-E}{\emph{Nucl. Phys. B}
  {\bfseries 387} (1992) 3}.

\bibitem{Duke:1982}
D.W.~Duke and J.F.~Owens, \emph{Quantum-chromodynamic corrections to
  deep-inelastic compton scattering},
  \href{https://doi.org/10.1103/PhysRevD.26.1600}{\emph{Phys. Rev. D}
  {\bfseries 26} (1982) 1600}.

\bibitem{Hekhorn:2019}
F.~Hekhorn, \emph{{Next-to-Leading Order QCD Corrections to Heavy-Flavour
  Production in Neutral Current DIS}}, Ph.D. thesis, {University of
  T\"ubingen}, 2019.
\newblock \href{https://arxiv.org/abs/1910.01536}{{\ttfamily 1910.01536}}.
\newblock 10.15496/publikation-34811.

\bibitem{Anderle:2016}
D.~Anderle, D.~de~Florian and Y.~Rotstein~Habarnau, \emph{{Towards
  semi-inclusive deep inelastic scattering at next-to-next-to-leading order}},
  \href{https://doi.org/10.1103/PhysRevD.95.034027}{\emph{Phys. Rev. D}
  {\bfseries 95} (2017) 034027}
  [\href{https://arxiv.org/abs/1612.01293}{{\ttfamily 1612.01293}}].

\bibitem{Wang:2019}
B.~Wang, J.~Gonzalez-Hernandez, T.~Rogers and N.~Sato, \emph{{Large Transverse
  Momentum in Semi-Inclusive Deeply Inelastic Scattering Beyond Lowest Order}},
  \href{https://doi.org/10.1103/PhysRevD.99.094029}{\emph{Phys. Rev. D}
  {\bfseries 99} (2019) 094029}
  [\href{https://arxiv.org/abs/1903.01529}{{\ttfamily 1903.01529}}].

\bibitem{Gordon:1993}
L.~Gordon and W.~Vogelsang, \emph{{Polarized and unpolarized prompt photon
  production beyond the leading order}},
  \href{https://doi.org/10.1103/PhysRevD.48.3136}{\emph{Phys. Rev. D}
  {\bfseries 48} (1993) 3136}.

\bibitem{Coriano:1996}
C.~Coriano and L.~Gordon, \emph{Polarized and unpolarized double prompt photon
  production in next-to-leading order qcd},
  \href{https://doi.org/https://doi.org/10.1016/0550-3213(96)00139-3}{\emph{Nuclear
  Physics B} {\bfseries 469} (1996) 202}.

\bibitem{Rein:2024}
D.~Rein, M.~Schlegel and W.~Vogelsang, \emph{Probing the polarized photon
  content of the proton in $ep$ collisions at the eic},
  \href{https://doi.org/10.1103/PhysRevD.110.014041}{\emph{Phys. Rev. D}
  {\bfseries 110} (2024) 014041}
  [\href{https://arxiv.org/abs/2405.04232}{{\ttfamily 2405.04232}}].

\bibitem{Ellis:1980}
R.K.~Ellis, M.A.~Furman, H.E.~Haber and I.~Hinchliffe, \emph{{Large Corrections
  to High p(T) Hadron-Hadron Scattering in QCD}},
  \href{https://doi.org/10.1016/0550-3213(80)90010-3}{\emph{Nucl. Phys. B}
  {\bfseries 173} (1980) 397}.

\bibitem{Schlegel:2012}
M.~Schlegel, \emph{{Partonic description of the transverse target single-spin
  asymmetry in inclusive deep-inelastic scattering}},
  \href{https://doi.org/10.1103/PhysRevD.87.034006}{\emph{Phys. Rev. D}
  {\bfseries 87} (2013) 034006}
  [\href{https://arxiv.org/abs/1211.3579}{{\ttfamily 1211.3579}}].

\bibitem{Ringer:2015}
F.~Ringer and W.~Vogelsang, \emph{{Single-Spin Asymmetries in W Boson
  Production at Next-to-Leading Order}},
  \href{https://doi.org/10.1103/PhysRevD.91.094033}{\emph{Phys. Rev. D}
  {\bfseries 91} (2015) 094033}
  [\href{https://arxiv.org/abs/1503.07052}{{\ttfamily 1503.07052}}].

\bibitem{tHooft:1972}
G.~{'t Hooft} and M.~Veltman, \emph{Regularization and renormalization of gauge
  fields},
  \href{https://doi.org/https://doi.org/10.1016/0550-3213(72)90279-9}{\emph{Nucl.
  Phys. B} {\bfseries 44} (1972) 189 }.

\bibitem{Bollini:1972}
{C.\,G. Bollini and J.\,J. Giambiagi}, \emph{{Dimensional Renormalization: The
  Number of Dimensions as a Regularizing Parameter}},
  \href{https://doi.org/10.1007/BF02895558}{\emph{{Nuovo Cim. B}} {\bfseries
  12} (1972) 20}.

\bibitem{Salvatori:2024}
G.~Salvatori, \emph{{The Tropical Geometry of Subtraction Schemes}},
  \href{https://arxiv.org/abs/2406.14606}{{\ttfamily 2406.14606}}.

\bibitem{Kotikov:1990}
A.V.~Kotikov, \emph{{Differential equations method: New technique for massive
  Feynman diagrams calculation}},
  \href{https://doi.org/10.1016/0370-2693(91)90413-K}{\emph{Phys. Lett. B}
  {\bfseries 254} (1991) 158}.

\bibitem{Remiddi:1997}
E.~Remiddi, \emph{{Differential equations for Feynman graph amplitudes}},
  \href{https://doi.org/10.1007/BF03185566}{\emph{Nuovo Cim. A} {\bfseries 110}
  (1997) 1435} [\href{https://arxiv.org/abs/hep-th/9711188}{{\ttfamily
  hep-th/9711188}}].

\bibitem{Gehrmann:1999}
T.~Gehrmann and E.~Remiddi, \emph{{Differential equations for two-loop
  four-point functions}},
  \href{https://doi.org/10.1016/S0550-3213(00)00223-6}{\emph{Nucl. Phys. B}
  {\bfseries 580} (2000) 485}
  [\href{https://arxiv.org/abs/hep-ph/9912329}{{\ttfamily hep-ph/9912329}}].

\bibitem{Henn:2013}
J.M.~Henn, \emph{{Multiloop integrals in dimensional regularization made
  simple}}, \href{https://doi.org/10.1103/PhysRevLett.110.251601}{\emph{Phys.
  Rev. Lett.} {\bfseries 110} (2013) 251601}
  [\href{https://arxiv.org/abs/1304.1806}{{\ttfamily 1304.1806}}].

\bibitem{Henn:2014}
J.M.~Henn, \emph{{Lectures on differential equations for Feynman integrals}},
  \href{https://doi.org/10.1088/1751-8113/48/15/153001}{\emph{J. Phys. A}
  {\bfseries 48} (2015) 153001}
  [\href{https://arxiv.org/abs/1412.2296}{{\ttfamily 1412.2296}}].

\bibitem{Badger:2023}
S.~Badger, J.~Henn, J.C.~Plefka and S.~Zoia, \emph{{Scattering Amplitudes in
  Quantum Field Theory}},
  \href{https://doi.org/10.1007/978-3-031-46987-9}{\emph{Lect. Notes Phys.}
  {\bfseries 1021} (2024) pp.}
  [\href{https://arxiv.org/abs/2306.05976}{{\ttfamily 2306.05976}}].

\bibitem{Tarasov:1996}
O.V.~Tarasov, \emph{Connection between feynman integrals having different
  values of the space-time dimension},
  \href{https://doi.org/10.1103/PhysRevD.54.6479}{\emph{Phys. Rev. D}
  {\bfseries 54} (1996) 6479}.

\bibitem{Lee:2012}
R.N.~Lee and V.A.~Smirnov, \emph{The dimensional recurrence and analyticity
  method for multicomponent master integrals: using unitarity cuts to construct
  homogeneous solutions},
  \href{https://doi.org/10.1007/jhep12(2012)104}{\emph{Journal of High Energy
  Physics} {\bfseries 2012} (2012) }.

\bibitem{Binosi:2003}
D.~Binosi and L.~Theu\ss{}l, \emph{{JaxoDraw: A Graphical user interface for
  drawing Feynman diagrams}},
  \href{https://doi.org/10.1016/j.cpc.2004.05.001}{\emph{Comput. Phys. Commun.}
  {\bfseries 161} (2004) 76}
  [\href{https://arxiv.org/abs/hep-ph/0309015}{{\ttfamily hep-ph/0309015}}].

\bibitem{Anastasiou:2002}
C.~Anastasiou and K.~Melnikov, \emph{Higgs boson production at hadron colliders
  in nnlo qcd},
  \href{https://doi.org/10.1016/s0550-3213(02)00837-4}{\emph{Nuclear Physics B}
  {\bfseries 646} (2002) 220–256}.

\bibitem{Anastasiou:2003}
C.~Anastasiou, L.~Dixon, K.~Melnikov and F.~Petriello, \emph{Dilepton rapidity
  distribution in the drell-yan process at next-to-next-to-leading order in
  qcd}, \href{https://doi.org/10.1103/PhysRevLett.91.182002}{\emph{Phys. Rev.
  Lett.} {\bfseries 91} (2003) 182002}.

\bibitem{Anastasiou:2004}
C.~Anastasiou, L.~Dixon, K.~Melnikov and F.~Petriello, \emph{High-precision qcd
  at hadron colliders: Electroweak gauge boson rapidity distributions at
  next-to-next-to leading order},
  \href{https://doi.org/10.1103/PhysRevD.69.094008}{\emph{Phys. Rev. D}
  {\bfseries 69} (2004) 094008}.

\bibitem{Smirnov:2019}
A.V.~Smirnov and F.S.~Chukharev, \emph{{FIRE6: Feynman Integral REduction with
  modular arithmetic}},
  \href{https://doi.org/10.1016/j.cpc.2019.106877}{\emph{Comput. Phys. Commun.}
  {\bfseries 247} (2020) 106877}
  [\href{https://arxiv.org/abs/1901.07808}{{\ttfamily 1901.07808}}].

\bibitem{Klappert:2020}
J.~Klappert, F.~Lange, P.~Maierh\"ofer and J.~Usovitsch, \emph{{Integral
  reduction with Kira 2.0 and finite field methods}},
  \href{https://doi.org/10.1016/j.cpc.2021.108024}{\emph{Comput. Phys. Commun.}
  {\bfseries 266} (2021) 108024}
  [\href{https://arxiv.org/abs/2008.06494}{{\ttfamily 2008.06494}}].

\bibitem{Lee:2013}
R.N.~Lee, \emph{{LiteRed 1.4: a powerful tool for reduction of multiloop
  integrals}}, \href{https://doi.org/10.1088/1742-6596/523/1/012059}{\emph{J.
  Phys. Conf. Ser.} {\bfseries 523} (2014) 012059}
  [\href{https://arxiv.org/abs/1310.1145}{{\ttfamily 1310.1145}}].

\bibitem{Liu:2024}
Z.L.~Liu and P.F.~Monni, \emph{{The two-loop fully differential soft function
  for $Q\bar{Q}V$ production at lepton colliders}},
  \href{https://arxiv.org/abs/2411.13466}{{\ttfamily 2411.13466}}.

\bibitem{Baranowski:2022}
D.~Baranowski, M.~Delto, K.~Melnikov and C.-Y.~Wang, \emph{On phase-space
  integrals with heaviside functions},
  \href{https://doi.org/10.1007/jhep02(2022)081}{\emph{Journal of High Energy
  Physics} {\bfseries 2022} (2022) }.

\bibitem{Laporta:2000}
S.~Laporta, \emph{{High-precision calculation of multiloop Feynman integrals by
  difference equations}},
  \href{https://doi.org/10.1142/S0217751X00002159}{\emph{Int. J. Mod. Phys. A}
  {\bfseries 15} (2000) 5087}
  [\href{https://arxiv.org/abs/hep-ph/0102033}{{\ttfamily hep-ph/0102033}}].

\bibitem{tHooft:1978}
G.~'t~Hooft and M.J.G.~Veltman, \emph{{Scalar One Loop Integrals}},
  \href{https://doi.org/10.1016/0550-3213(79)90605-9}{\emph{Nucl. Phys. B}
  {\bfseries 153} (1979) 365}.

\bibitem{Besier:2018}
M.~Besier, D.~Van~Straten and S.~Weinzierl, \emph{{Rationalizing roots: an
  algorithmic approach}},
  \href{https://doi.org/10.4310/CNTP.2019.v13.n2.a1}{\emph{Commun. Num. Theor.
  Phys.} {\bfseries 13} (2019) 253}
  [\href{https://arxiv.org/abs/1809.10983}{{\ttfamily 1809.10983}}].

\bibitem{Euler:1781}
L.~Euler, \emph{De mensura angulorum solidorum}, {\emph{Acta Academiae
  Scientiarum Imperialis Petropolitanae} (1781) 31}.

\bibitem{Eriksson:1990}
F.~Eriksson, \emph{On the measure of solid angles},
  \href{https://doi.org/10.1080/0025570X.1990.11977515}{\emph{Mathematics
  Magazine} {\bfseries 63} (1990) 184}.

\bibitem{Ahmed:2024}
T.~Ahmed, S.M.~Hasan and A.~Rapakoulias, \emph{{Phase-space integrals through
  Mellin-Barnes representation}},
  \href{https://arxiv.org/abs/2410.18886}{{\ttfamily 2410.18886}}.

\bibitem{Lewin:1981}
{L. Lewin}, \emph{{Polylogarithms and Associated Functions}}, {Elsevier}
  (1981).

\bibitem{Clausen:1832}
T.~Clausen, \emph{{\"Uber die Funktion $\sin\phi+\frac{1}{2^2}\sin
  2\phi+\frac{1}{3^2}\sin 3\phi+$etc.}}, {\emph{Journal für die reine und
  angewandte Mathematik} {\bfseries 8} (1832) 298}.

\bibitem{Wood:1968}
V.E.~Wood, \emph{{Efficient calculation of Clausen’s integral}}, {\emph{Math.
  Comp.} {\bfseries 22} (1968) 883}.

\bibitem{Kolbig:1995}
K.S.~Kolbig, \emph{{Chebyshev coefficients for the Clausen function
  $\mathrm{Cl}_2(x)$}},
  \href{https://doi.org/10.1016/0377-0427(95)00150-6}{\emph{J. Comput. Appl.
  Math.} {\bfseries 64} (1995) 295}.

\bibitem{Wu:2010}
J.~Wu, X.~Zhang and D.~Liu, \emph{{An efficient calculation of the Clausen
  functions $\mathrm{Cl}_n (\theta)(n\geq 2)$}},
  \href{https://doi.org/10.1007/s10543-009-0246-8}{\emph{BIT Numerical
  Mathematics} {\bfseries 50} (2010) 193}.

\bibitem{Zagier:2007}
D.~Zagier, \emph{The dilogarithm function},  in \emph{{Frontiers in Number
  Theory, Physics, and Geometry II: On Conformal Field Theories, Discrete
  Groups and Renormalization}}, pp.~3--65, Springer (2007),
  \href{{https://people.mpim-bonn.mpg.de/zagier/files/doi/10.1007/978-3-540-30308-4\_1/fulltext.pdf}}{{https://people.mpim-bonn.mpg.de/zagier/files/doi/10.1007/978-3-540-30308-4\_1/fulltext.pdf}}.

\bibitem{Lu:1992}
H.J.~Lu and C.A.~Perez, \emph{{Massless one loop scalar three point integral
  and associated Clausen, Glaisher and L functions}}, .

\bibitem{Wagner:1996}
P.~Wagner, \emph{A volume formula for asymptotic hyperbolic tetrahedra with an
  application to quantum field theory},
  \href{https://doi.org/https://doi.org/10.1016/S0019-3577(97)89138-0}{\emph{Indagationes
  Mathematicae} {\bfseries 7} (1996) 527}.

\bibitem{Davydychev:1997}
A.I.~Davydychev and R.~Delbourgo, \emph{{A Geometrical angle on Feynman
  integrals}}, \href{https://doi.org/10.1063/1.532513}{\emph{J. Math. Phys.}
  {\bfseries 39} (1998) 4299}
  [\href{https://arxiv.org/abs/hep-th/9709216}{{\ttfamily hep-th/9709216}}].

\bibitem{Broadhurst:1998}
D.J.~Broadhurst, \emph{{Solving differential equations for three loop diagrams:
  Relation to hyperbolic geometry and knot theory}},
  \href{https://arxiv.org/abs/hep-th/9806174}{{\ttfamily hep-th/9806174}}.

\bibitem{Broadhurst:1999}
D.J.~Broadhurst, \emph{{Massive three - loop Feynman diagrams reducible to SC*
  primitives of algebras of the sixth root of unity}},
  \href{https://doi.org/10.1007/s100529900935}{\emph{Eur. Phys. J. C}
  {\bfseries 8} (1999) 311}
  [\href{https://arxiv.org/abs/hep-th/9803091}{{\ttfamily hep-th/9803091}}].

\bibitem{Davydychev:2000}
A.I.~Davydychev and M.Y.~Kalmykov, \emph{{New results for the epsilon expansion
  of certain one, two and three loop Feynman diagrams}},
  \href{https://doi.org/10.1016/S0550-3213(01)00095-5}{\emph{Nucl. Phys. B}
  {\bfseries 605} (2001) 266}
  [\href{https://arxiv.org/abs/hep-th/0012189}{{\ttfamily hep-th/0012189}}].

\bibitem{Davydychev:2003}
A.I.~Davydychev and M.Y.~Kalmykov, \emph{{Massive Feynman diagrams and inverse
  binomial sums}},
  \href{https://doi.org/10.1016/j.nuclphysb.2004.08.020}{\emph{Nucl. Phys. B}
  {\bfseries 699} (2004) 3}
  [\href{https://arxiv.org/abs/hep-th/0303162}{{\ttfamily hep-th/0303162}}].

\bibitem{Czakon:2004}
M.~Czakon, \emph{{The Four-loop QCD beta-function and anomalous dimensions}},
  \href{https://doi.org/10.1016/j.nuclphysb.2005.01.012}{\emph{Nucl. Phys. B}
  {\bfseries 710} (2005) 485}
  [\href{https://arxiv.org/abs/hep-ph/0411261}{{\ttfamily hep-ph/0411261}}].

\bibitem{Binger:2006}
M.~Binger and S.J.~Brodsky, \emph{{The Form-factors of the gauge-invariant
  three-gluon vertex}},
  \href{https://doi.org/10.1103/PhysRevD.74.054016}{\emph{Phys. Rev. D}
  {\bfseries 74} (2006) 054016}
  [\href{https://arxiv.org/abs/hep-ph/0602199}{{\ttfamily hep-ph/0602199}}].

\bibitem{Adams:2015}
L.~Adams, C.~Bogner and S.~Weinzierl, \emph{{The two-loop sunrise integral
  around four space-time dimensions and generalisations of the Clausen and
  Glaisher functions towards the elliptic case}},
  \href{https://doi.org/10.1063/1.4926985}{\emph{J. Math. Phys.} {\bfseries 56}
  (2015) 072303} [\href{https://arxiv.org/abs/1504.03255}{{\ttfamily
  1504.03255}}].

\bibitem{Chicherin:2020}
D.~Chicherin and V.~Sotnikov, \emph{{Pentagon Functions for Scattering of Five
  Massless Particles}},
  \href{https://doi.org/10.1007/JHEP12(2020)167}{\emph{JHEP} {\bfseries 20}
  (2020) 167} [\href{https://arxiv.org/abs/2009.07803}{{\ttfamily
  2009.07803}}].

\bibitem{Abreu:2022}
S.~Abreu, M.~Becchetti, C.~Duhr and M.A.~Ozcelik, \emph{{Two-loop master
  integrals for pseudo-scalar quarkonium and leptonium production and decay}},
  \href{https://doi.org/10.1007/JHEP09(2022)194}{\emph{JHEP} {\bfseries 09}
  (2022) 194} [\href{https://arxiv.org/abs/2206.03848}{{\ttfamily
  2206.03848}}].

\end{thebibliography}\endgroup
\bibliographystyle{JHEP}
\end{document}